\RequirePackage[hyphens]{url}

\documentclass[5p,twocolumn,sort&compress]{elsarticle}
\makeatletter
\def\ps@pprintTitle{%
 \let\@oddhead\@empty
 \let\@evenhead\@empty
 \def\@oddfoot{\centerline{\thepage}}%
 \let\@evenfoot\@oddfoot}
\makeatother

\usepackage{epsfig,endnotes}
\usepackage{paralist}
\usepackage{graphicx}
\usepackage[utf8]{inputenc}
\usepackage{listings}
\usepackage{ntheorem}
\theorembodyfont{\small}

\usepackage{url}
\usepackage{framed}
\usepackage{lmodern}
\usepackage[T1]{fontenc}
\usepackage{inputenc}
\usepackage{multirow}
\usepackage{multicol}
\usepackage{tabularx}
\usepackage{booktabs}
\usepackage{algorithm}
\usepackage{algorithmicx}
\usepackage{algpseudocode}
\usepackage{enumitem}
\usepackage{color}
\usepackage{ifthen}
\newboolean{showcomments}
\setboolean{showcomments}{true}
\ifthenelse{\boolean{showcomments}}
 { \newcommand{\mynote}[2]{
      \fbox{\bfseries\sffamily\scriptsize#1}
        {\small$\blacktriangleright$\textsf{\textcolor{red}{{\em #2}\bf }}$\blacktriangleleft$}}}
        { \newcommand{\mynote}[2]{}}

\newboolean{hidetext}
\setboolean{hidetext}{true}
\ifthenelse{\boolean{hidetext}}
  { \newcommand{\hide}[1]{}}
  { \newcommand{\hide}[1]{#1}}

\definecolor{lightgray}{gray}{.50}

\usepackage{cellspace}
\usepackage{makecell}

\newcommand{\eg}{\textit{e.g., }}
\newcommand{\ie}{\textit{i.e., }}
\newcommand{\cf}{\textit{cf. }}

\newcommand{\SYS}[0]{\textsc{GreyCat}}

\makeatletter
\newcommand{\newalgname}[1]{
  \renewcommand{\ALG@name}{#1}
}
\newalgname{Algorithm}
\makeatother

\begin{document}



\title{\SYS: Efficient What-If Analytics for Data in Motion at Scale}

\author[ul]{Thomas~Hartmann}
\ead{thomas.hartmann@uni.lu}

\author[dt]{Francois~Fouquet}
\ead{francois.fouquet@uni.lu}

\author[dt]{Assaad~Moawad}
\ead{assaad.moawad@datathings.lu}

\author[ui,i]{Romain~Rouvoy}
\ead{romain.rouvoy@univ-lille.fr}

\author[ul]{Yves~Le~Traon}
\ead{yves.letraon@uni.lu}

\address[ul]{University of Luxembourg, Luxembourg}
\address[dt]{DataThings, Luxembourg}
\address[ui]{University of Lille / Inria, France}
\address[i]{Institut Universitaire de France (IUF)}




\begin{abstract}
Over the last few years, data analytics shifted from a descriptive era, confined to the explanation of past events, to the emergence of predictive techniques.
Nonetheless, existing predictive techniques still fail to effectively explore alternative futures, which continuously diverge from current situations when exploring the effects of \emph{what-if} decisions.
Enabling prescriptive analytics therefore calls for the design of scalable systems that can cope with the complexity and the diversity of underlying data models.
In this article, we address this challenge by combining graphs and time series within a scalable storage system that can organize a massive amount of unstructured and continuously changing data into multi-dimensional data models, called \emph{Many-Worlds Graphs}.
We demonstrate that our open source implementation, \SYS{}, can efficiently fork and update thousands of parallel worlds composed of millions of timestamped nodes, such as \emph{what-if} exploration.
\end{abstract}

\maketitle

\section{Introduction}\label{sec:introduction}

The data deluge raised by large-scale distributed systems has called for scalable analytics platforms in order to guide decisions in critical cyber-physical infrastructures, such as smart grids~\cite{Kathiravelu:2015:CMP:2836127.2836132}.
In this domain, \emph{predictive analytics} techniques, like sliding window analytics~\cite{bhatotia2014slider}, typically extract temporal models from current and past historical facts in order to make predictions about the future~\cite{Rusitschka:2013:AMR:2541596.2541601}.
However, taking appropriate decisions rather requires \emph{prescriptive analytics} in order to explore the impact of current and future actions on the underlying system, better known as \emph{what-if analysis}~\cite{HaasVLDB11}.
More specifically, what-if analysis is a powerful primitive to plan an optimal sequence of actions that leads to a desired target state of the underlying system.
Reaching this optimization objective implies to cover all potential decision timepoints and applicable actions, thus inevitably yielding to a combinatorial explosion of alternative scenarios.
In addition to that, this decision process has to deal with the continuous updates of states as time keeps flowing along.

\begin{figure}
	\centering
	\includegraphics[width=\linewidth]{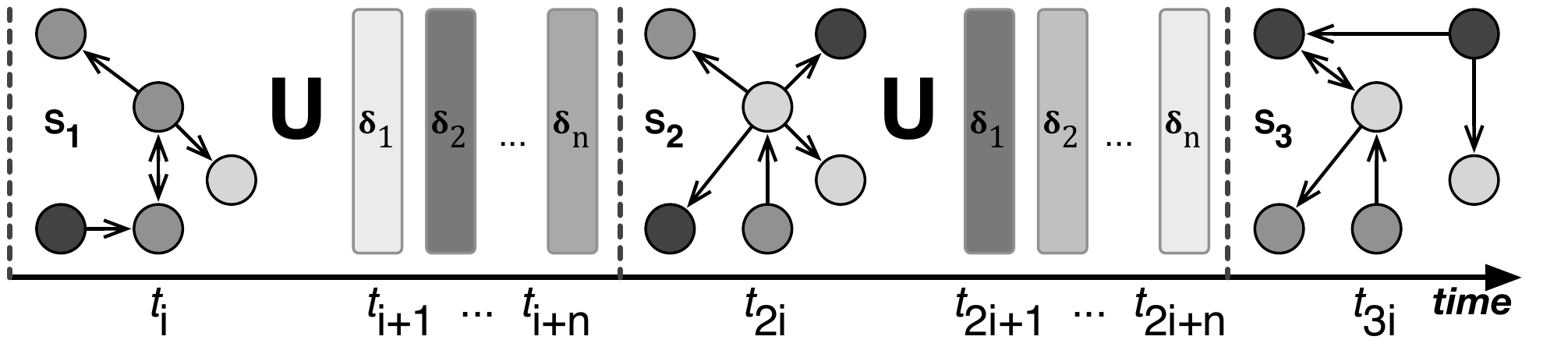}
	\caption{Snapshots ($S_i$) and deltas ($\delta_n$) of a time-evolving graph}
	\label{fig:snapshots}
\end{figure}

Graphs are increasingly being used to structure and analyze such complex data~\cite{Gonzalez:2012:PDG:2387880.2387883,Low:2012:DGF:2212351.2212354,Malewicz:2010:PSL:1807167.1807184}.
However, most of graph representations only reflect a snapshot at a given time, while reflected data keeps changing as the systems evolve.
Understanding temporal characteristics of time-evolving graphs therefore attracts increasing attention from research communities~\cite{Leskovec:2005:GOT:1081870.1081893}---\eg in the domains of social networks, smart mobility, or smart grids~\cite{hartmannSeke14}.
Yet, state-of-the-art approaches fail to provide a scalable solution to effectively support time in graphs.
In particular, existing approaches represent time-evolving graphs as sequences of full-graph snapshots~\cite{Iyer:2016:TGP:2960414.2960419}, or they use a combination of snapshots and deltas~\cite{Han:2014:CGE:2592798.2592799}, which requires to reconstruct a graph for a given time, as depicted in Figure~\ref{fig:snapshots}.

However, full-graph snapshots tend to be expensive in terms of memory requirements, both on disk and in-memory.
This overhead becomes even worse when data from several snapshots need to be correlated, which is the case for most of advanced analytics~\cite{hartmannSeke14, Iyer:2016:TGP:2960414.2960419, Miao:2015:ISS:2809503.2700302}.
Another challenging issue related to snapshots relates to the snapshotting frequency: regardless of changes, for any change in the graph, or only for the major changes, which results in a tradeoff between duplicating data and feeding analytics with up-to-date metrics.
This is crucial when data evolves rapidly and at different paces for different elements in the graph, like it is for example the case with sensor data in domains like the \emph{Internet of Things} (IoT) or \emph{Cyber-Physical Systems} (CPS)~\cite{hartmannSeke14}.
An alternative to snapshotting consists in combining graphs with time series databases~\cite{Lin:2003:SRT:882082.882086}, by mapping individual nodes to time series.
However, this becomes quickly limited when large parts of the graph evolve over time, inducing multiple time queries to explore the graph.
Moreover, the description and the evolution of relationships among the nodes of the graph are rather hard to model within a time series database.

These challenges are even exacerbated when it comes to what-if analysis on top of such time-evolving graphs.

Therefore, in this article, we propose to adopt the theory of \emph{many-worlds interpretation}~\cite{everett1957relative}, where every single action can be interpreted as a divergence point, forking an alternative, independent world.
In particular, we introduce the concept of \emph{Many-Worlds Graphs} (MWG) as a versatile and scalable analytics data model supporting the evaluation of hundreds or even thousands of alternative actions on temporal graphs in parallel.
MWG extends state-of-the-art graph analytics~\cite{low2014graphlab,Malewicz:2010:PSL:1807167.1807184,xin2013graphx}, which are commonly used to organize massive amounts of unstructured data~\cite{Malewicz:2010:PSL:1807167.1807184,miller2013graph}.
Beyond the inclusion of temporal aspects within graphs~\cite{Bahmani:2012:PEG:2339530.2339539,Cattuto:2013:TSN:2484425.2484442,Iyer:2016:TGP:2960414.2960419,khurana2015storing}, MWG proposes an efficient exploration of many independently evolving worlds to support the requirements of what-if analysis.
The main contribution of this paper is a novel graph data model to support large-scale what-if analysis on time-evolving graphs.
Related topics like graph processing, traversing, fault tolerance, and distribution are described where necessary.

We demonstrate that \SYS{}, our implementation of MWG, can efficiently explore hundreds of thousands of independent worlds in parallel and we assess this capability on a real-world smart grid's workload.
GreyCat is open source and available at GitHub\footnote{https://github.com/datathings/greycat}.
MWG and \SYS{} refer to Everett's~\cite{everett1957relative} many-world interpretation, illustrated by Schrödinger's cat~\cite{schrodinger1935present}.

The remainder of this article is organized as follows.
Section~\ref{sec:motivation} introduces a real smart grid case study, as a motivation of this research.
Sections~\ref{sec:overview} and~\ref{sec:middleware_implementation} introduce the main concepts of MWG and their implementation.
We thoroughly evaluate \SYS{} in Section~\ref{sec:evaluation}.
The related work is discussed in Section~\ref{sec:related} before concluding in Section~\ref{sec:conclusion}.

\section{Motivating Case Study}\label{sec:motivation}

Intelligent load management is a critical challenge for electricity utility companies~\cite{1507024,5357331}.
These companies are expected to avoid overload situations in electricity \emph{cables} by balancing the load.
The electric load in cables depends on the current and historical consumption of customers connected to a given cable within the system topology---\emph{i.e.}, how cables are connected to each other and to power substations.

The underlying topology can be changed by opening/closing so-called \emph{fuses} at substations.
This results in connecting/disconnecting households to different power substations, therefore impacting the electricity flow within the grid.
As all of this (consumptions, decisions) changes over time, the idea behind prescriptive analytics is to continuously simulate the expected load for different topologies (what-if scenarios) with the goal to find an ``optimal'' one---\emph{i.e.}, where the load in all cables is the best balanced.
Smart grids are very-large-scale systems, connecting hundreds of thousands or even hundreds of millions of nodes (customers).
Furthermore, most data in the context of smart grids is temporal, \emph{i.e.}, it keeps changing over time, from consumption reports to the topology structure.
This makes the simulation of different what-if scenarios very challenging and, in addition, it requires to take the temporal dimension, \emph{i.e.,} data history---into account.
Moreover, many different topologies are possible, which can easily lead to thousands of different scenarios.

To anticipate potential overload situations, alternative topologies need to be explored \emph{a priori}---\emph{i.e.}, before the problem actually occurs.
The estimation of the electric load depends, aside from the topology, on the consumption data of customers.
In the context of a smart grid, this data is measured by \emph{smart meters}, which are installed at customers' homes, and regularly report to utility companies, (\emph{e.g.}, every 15 minutes~\cite{DBLP:conf/smartgridcomm/0001MFRMKT15}).
One can compute the electric load based on the profiles of customers' consumption behavior.
These profiles are built using online machine learning algorithms, such as the ones introduced in~\cite{DBLP:conf/smartgridcomm/0001MFRMKT15}.
However, the huge amount of consumption data quickly leads to millions of values per customer, and efficiently analyzing such large historical datasets is challenging.
The temporal dimension of data often results in inefficient data querying and iteration operations to find the requested data.
While this issue has been extensively discussed by the database community in the 80s and 90s~\cite{clifford1983formal,segev1987logical}, this topic is gaining popularity again with the advent of \textit{time series} databases for the IoT, like InfluxDB~\cite{influxdb}.
Time series can be seen as a special kind of temporal data, which is defined as a sequence of timestamped data points, and is used to store data like ocean tides, stock values, and weather data.
It is important to note that in time series, data is ``flat'', \emph{i.e.}, time series only contain primitive tuples, like raw measurements.
However, they are not able to capture complex data structures and their relationships like, for example, the evolution of a smart grid topology.
Therefore, time series analysis is not sufficient to explore complex what-if analysis and prescriptive analytics.
On the other side, graph-based storage solutions (\emph{e.g.}, Neo4J~\cite{neo4j}), as well as graph processing frameworks, despite some attempts to represent time dependent graphs~\cite{Cattuto:2013:TSN:2484425.2484442,Cheng:2012:KTP:2168836.2168846,Han:2014:CGE:2592798.2592799,Iyer:2016:TGP:2960414.2960419}, are insufficiently addressing continuously changing data in their model: either failing to navigate through alternative versions of a given graph, or inefficiently covering this issue by generating distinct snapshots of the graph.
When we speak in this paper about temporal data, we refer to this fully temporal aspect and not about flat time series'.
Most importantly, none of these solutions supports the large-scale exploration of different alternatives.

These limitations motivates our work on MWG in order to enable large-scale what-if analysis for prescriptive analytics~\cite{HaasVLDB11}.
Haas \emph{et al.}~\cite{HaasVLDB11}, also supports the need for large-scale what-if analysis in many other domains, \emph{e.g.,} weather prediction.
Beyond the specificities of smart grids, we introduce the concept of \emph{Many-World Graphs} as a scalable model to explore alternative scenarios in the context of what-if analysis.

\section{Introducing Many-World Graphs}\label{sec:overview}
\subsection{Key Concepts}
This paper introduces the notion of \emph{Many-World Graphs} (MWG), which are directed, attributed hypergraphs which structure and properties can evolve along time and parallel worlds.
In particular, MWG build on the following core concepts:

\begin{compactdesc}
	\item[\textsf{Timepoint}] is an event, encoded as a timestamp;
	\item[\textsf{World}] is a parallel universe, used as an identifier;
	\item[\textsf{Node}] reflects a domain-specific concept, which exists across \textsf{worlds}, and is used as an identifier;
	\item[\textsf{State}] is a \textsf{node}'s value for a given \textsf{world} and \textsf{timepoint}, including attributes and relationships;
	\item[\textsf{Timeline}] is a sequence of states for a given \textsf{node} and a given \textsf{world}.
\end{compactdesc}

\begin{figure}
 	\centering
 	\includegraphics[width=.65\linewidth]{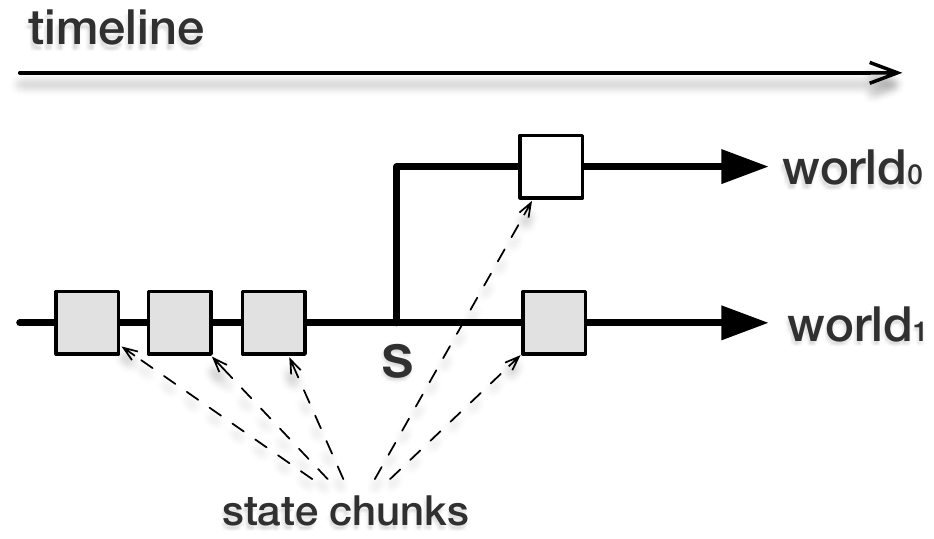}
 	\caption{States of a node in two worlds}
 	\label{fig:simple_state_chunk}
\end{figure}

Depending on the \textsf{timepoint} ($t$) and \textsf{world} ($w$), different \textsf{states} can be fetched from a given \textsf{node} ($n$).
This is illustrated in Figure~\ref{fig:simple_state_chunk}.
Therefore, states are organized into \textsf{chunks} ($c$), which are uniquely mapped from any viewpoint:

\bigskip
\qquad $\langle{n,t,w}\rangle: read(n, t, w) \mapsto c_t$.
\bigskip

We associate each state chunk with a timepoint ($c_t$) and we define a \textsf{timeline} ($t_{n,w} = [c_{0}, \ldots, c_{n}]$) as an ordered sequence of chunks belonging to a given node ($n$) from a given world ($w$).
Alternative state chunks in different worlds therefore form alternative \textsf{timelines}.
Hence, a resolution function $read$ returns a chunk ($c_t$) for an input viewpoint as the ``closest'' state chunk in the \textsf{timeline}.

Therefore, when a MWG is explored, state chunks of every node have to be resolved according to an input world and timepoint.
The storage and processing of MWG made of billions of nodes cannot be done in memory, thus requiring to efficiently store and retrieve chunks from a persistent data store.
For this purpose, we decompose state chunks into keys and values to leverage existing key/value stores to persist the data on disk.
The mapping of nodes to state chunks (including attributes and references to other nodes) and their persistent storage is detailed in Section~\ref{sec:graph_serialization}.

\begin{figure}
 	\centering
 	\includegraphics[width=\linewidth]{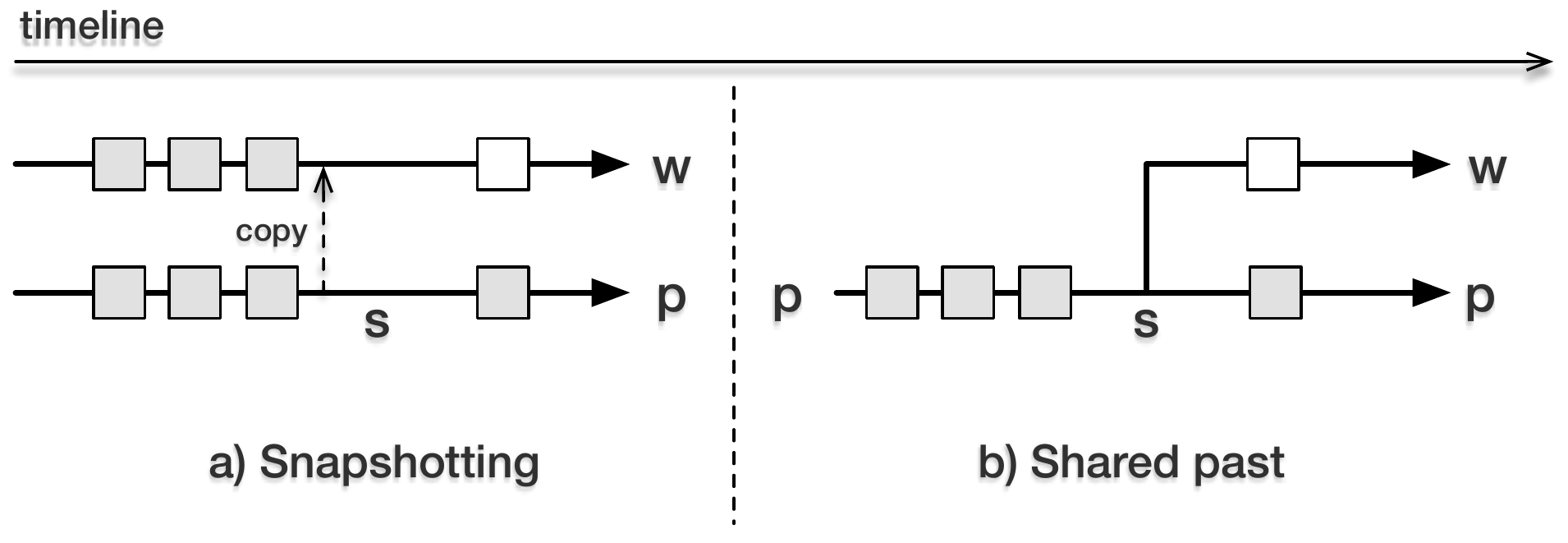}
 	\caption{Types of many-worlds.}
 	\label{fig:typesworlds}
 \end{figure}

While prescriptive analytics tends to explore new worlds along time, two techniques can be employed when forking worlds: \emph{snapshotting} and \emph{shared past} (cf. Figure~\ref{fig:typesworlds}).

\textbf{Snapshotting:} consists in copying all state chunks of all timepoints from a \emph{parent world} $p$ to the \emph{child world} $w$, thus leaving both worlds to evolve completely independently, from past to future.
Although this approach is simple, it is very inefficient in terms of time and storage to clone all state chunks of all historical records.


\textbf{Shared past:} we propose to adopt an alternative approach, which makes it unnecessary to copy past state chunks.
Instead, a new world $w$ is diverged from a parent $p$ at a point $s$ in time.
Before $t_s$, both worlds share the same past, thus resolving the same state chunks.
After $t_s$, world $w$ and $p$ \emph{co-evolve}, which means that each have their own timeline for $t_n \geq t_s$.
Therefore, both worlds share the same past before the divergent point (for $t<s$), but each evolves independently after the divergent point for $t \geq s$.

\subsection{Many-World Graph Semantics}\label{sec:manyworld_graph_semantic}
With MWG, we seek to efficiently organize and analyze data that can evolve along time and alternative worlds.
We define such a complex topology as a graph $G = N \times T \times W$, where $N$ is the set of nodes, $T$ the set of timepoints, and $W$ the set of worlds.
However, what-if analysis needs to explore many different actions, which usually does not affect all data in all worlds and all timepoints.
To address this combinatorial problem of world and timepoint alternatives, we define our MWG so that values of each node are resolved on-demand, based on a reference world and timepoint.
In this section, we formalize the semantics of our MWG by starting with a base graph definition, which we first extend with temporal semantics and then with the many-worlds semantics.

\subsection{Base Graph (BG)}\label{sec:graph_semantic:simple_graph_model}
A graph $G$ is commonly defined as an ordered pair $G =\{V, E\}$ consisting of a set $V$ of nodes and a set $E$ of edges.
In order to distinguish between nodes and their states, we define a different semantics.
First, we define a node as a conceptual identifier that is mapped to a ``state chunk''.
A state chunk contains the values of all attributes and edges that belong to a node.
Attributes are typed according to one of the following primitive types: \texttt{int}, \texttt{long}, \texttt{double}, \texttt{string}, \texttt{bool}, or \texttt{enumerations}.

The state chunk $c$ of a node $n$ is:

\medskip
\qquad $c_n=(A_n,R_n)$, where $A_n$ is the set of attribute values of node $n$ and $R_n$ is the set of relationship values from $n$ to other nodes.
\medskip

From now on, we refer to edges as directed relationships or simply as \emph{relationships}.
Unlike other graph models ({\em e.g.}, Neo4J~\cite{miller2013graph}), our model does not support edge attributes.
Nevertheless, any edge attribute can be easily modeled as an intermediate node within such graphs, without compromising the expressiveness and the efficiency.
Besides being simple, this also makes our graph data model similar to the object-oriented one, which today is the dominating data model of many modern programming languages, like Java, C\#, Scala and Swift. 
This straightforward mapping leverages the integration of MWG within an application layer.

Then, we introduce the function $read(n)$ to resolve the state chunk of a node $n$. 
It returns the state chunk of the node, which contains the relationships---or edges---to other nodes.
Thus, we define a base graph $BG$ as:

\medskip
\qquad $BG=\{read(n), \forall n \in N\}$.
\medskip

Unlike common graph definitions, our base graph is not statically defined, but dynamically created as the result of the evaluation of the $read(n)$ function over all nodes $n$.
Implicitly, all state chunks of all nodes are dynamically resolved and the graph aggregates the nodes accordingly to the relationships defined within the resolved state chunks. 
This definition forms the basis for the semantics of our proposed data model.

In this way, only the destination nodes need to be listed in the set, since all the directed edges start from the same node $n$, thus making it redundant to list the source node. 
For example, if we have: $c_n=\{\{att1\}, \{m, p\}\}$, where $m,p \in N$, this means that the node $n$ has one attribute and two relationships (one to node $m$ and another one to node $p$).
Two directed edges can be implicitly constructed: $n \rightarrow m$ and $n \rightarrow p$.
In the next sections, we overload the function read incrementally to integrate step by step the time and many-world semantic.

\subsection{Temporal Graph (TG)}\label{sec:graph_semantic:temporal_graph_model}
To extend our $BG$ with temporal semantics, we override the function $read(n)$ with a function $read(n,t)$, with $t \in T$.
$T$ is a totally ordered sequence of all possible timepoints: $\forall t_i, t_j \in T: t_i \leq t_j \lor t_j \leq t_i$. 

We also extend the state chunk with a temporal representation:

\smallskip
\qquad $c_{n, t}=(A_{n,t},R_{n,t})$, where $A_{n,t}$ and $R_{n,t}$ are the sets of resolved values of attributes and relationships, for the node $n$ at time $t$.
\smallskip

Then, we define the temporal graph as follows: 

\smallskip
\qquad $TG(t)=\{read(n, t), \forall n \in N\}, \forall t \in T$.
\smallskip

Every node of the $TG$ can evolve independently and, as timepoints can be compared, they naturally form a chronological order.
We define that every state chunk belonging to a node in a $TG$ is associated to a timepoint and can therefore be organized according to this chronological order in a sequence $TP \subseteq T$.
We call this ordered sequence of state chunks the \textit{timeline} of a node.
The timeline $tl$ of a node $n$ is defined as $tl_n=\{c_{n,t}, \forall t \in TP \subseteq T\}$.

The two core operations \texttt{insert} and \texttt{read} are defined as follows:

\medskip
\qquad $insert(c_{n,t}, n, t)\colon (c \times N \times T) \mapsto Void$, as the function that inserts a state chunk in the timeline of a node $n$, such as: $tl_n := tl_n \cup \{c_{n,t}\}$.
\medskip

The operation $read(n,t)$$\colon (N \times T) \mapsto c$, is the function that retrieves, from the timeline $tl_n$, and up until time $t$, the most recent version of the state chunk of $n$ which was inserted at timepoint $t_i$:

\[read(n,t)  =  \left\{ \begin{array}{ll}
c_{n,t_i} & \mbox{if $ (c_{n,t_i} \in tl_n) $} \\
          & \mbox{$ \wedge (t_i \in TP) \wedge (t_i<t)$} \\
          & \mbox{$\wedge  (\forall t_j \in TP \rightarrow t_j<t_i)$} \\
\emptyset & \mbox{otherwise}
\end{array}
\right. \]

\begin{figure}	
	\centering			
 	\includegraphics[width=.85\linewidth]{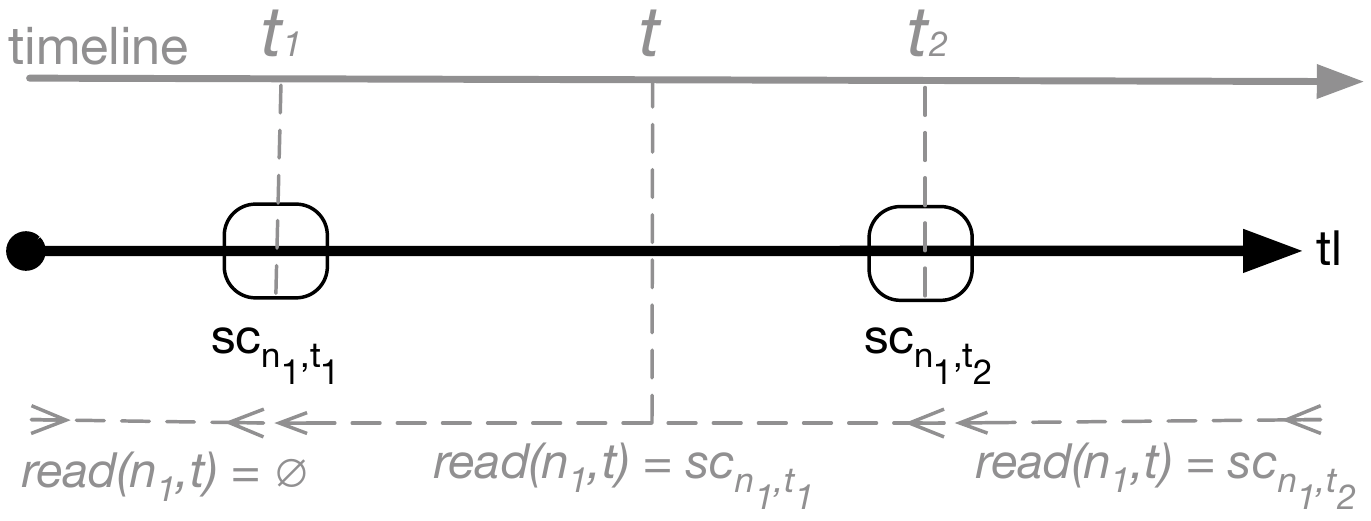}
 	\caption{TG node timeline} \label{fig:timeline}  
\end{figure}

Based on these definitions, although timestamps are discrete, they logically define intervals in which a state chunk can be considered as \textit{valid} within its timeline.
When executing $insert(c_{n_1,t_1}, n_{1}, t_{1})$ and $insert(c_{n_1,t_2}, n_{1}, t_{2})$, we insert 2 state chunks $c_{n_1,t_1}$ and $c_{n_1,t_2}$ for the same node $n_1$ at two different timepoints with $t_1<t_2$, we define that $c_{n_1,t_1}$ is valid in the open interval $[t_1,t_2[$, and $c_{n_1,t_2}$ is valid in $[t_2, +\infty[$.
Thus, an operation $read(n_{1}, t)$ resolves $\emptyset$ if $t<t_1$, $c_{n_1,t_1}$ when $t_1\leq t<t_2$, and $c_{n_1,t_2}$ if $t \geq t_2$ for the same node $n_1$.
The corresponding time validities are depicted in Figure~\ref{fig:timeline}.

Since state chunks with this semantics have temporal validities, relationships between nodes also have temporal validities.
This leads to \textit{temporal relationships} between TG nodes and forms a natural extension of relationships in the time dimension.
Once the time resolution returns the correct timepoint $t_i$, the temporal graph can be reduced to a base graph, therefore a TG for a particular $t$ can be seen as a base graph: $TG(t) \equiv BG_{t_i}$.

\subsection{Many-World Graph (MWG)}\label{sec:graph_semantic:manyworld_graph_model}
To extend the $TG$ with a many-world semantics, we refine the definition of the function $read(n,t)$ by considering, in addition to time, the different worlds.
The function $read(n,t,w)$, with $t \in T$ and $w \in W$, where $W$ is the set of all possible worlds, which resolves the state chunk of node $n$ at timepoint $t$ in world $w$.
In analogy to Section~\ref{sec:graph_semantic:temporal_graph_model}, the state chunk definition is extended as follows: 

\medskip
\qquad $c_{n,t,w} = (A_{n,t,w}, R_{n,t,w})$, where $A_{n,t,w}$ and $R_{n,t,w}$ are the sets of resolved values of attributes and relationships, for the node $n$ at time $t$, in world $w$.
\medskip

From this definition, a MWG is formalized as:

\medskip
\qquad $MWG(t,w) = \{ read(n,t,w), \forall n \in N\}, \forall (t,w) \in T \times W$, where $W$ is a partially ordered set of all possible worlds.
\medskip
 
The partial order $<$ on the set $W$ is defined by the \textbf{parent} ordering, with $(p < w) \equiv (p=parent(w))$.
Intuitively, the set $W$ is partially ordered by the generations of worlds.
However, worlds that are created from the same parent, or the worlds that are created from different parents, cannot be compared (in terms of order) to each other.
We define the first created world as the \textbf{root world}, with $parent(root)=\emptyset$. 
Then, all other worlds are created by diverging from the root world, or from any other existing world in the world map set $WM$ of our MWG.
The divergence function is defined as:

\medskip
\qquad $w=diverge(p)\colon World \mapsto World$, the function that creates world $w$ from the parent world $p$, with $p<w$ and $p \in WM \subseteq W$.
Upon divergence, we therefore obtain $WM:=WM \cup \{w\}$.
\medskip

According to this definition, we consider the world $w$ as the \textbf{child} of world $p$ and it is added to the world map of our MWG.
For the MWG, we define the \textbf{local timeline of a world and a node} as $ltl_{n,w} = \{ c_{n,t,w}, \forall t \in TP_{n,w}\}$,
with $TP_{n,w} \subseteq T$, which is the ordered subset of timepoints for node $n$ and world $w$.
As $TP_{n,w} $ is ordered, there exists a timepoint $s_{n,w}$, which is the smallest timepoint in $TP_{n,w}$, defined as $s_{n,w}\in TP_{n,w}, \forall t \in TP_{n,w}, s_{n,w}\leq t$. 
We call this timepoint a \textbf{divergent timepoint}---\emph{i.e.}, where the world $w$ starts to diverge from its parent $p$ for node $n$. 
Following the shared-past concept between a world and its parent (cf. Section~\ref{sec:overview}), we define the global timeline of a world per node as

\[\mbox{$tl(n,w)$} = \left\{
\begin{array}{lll}
	\mbox{$\emptyset$ if $w=\emptyset$} \\
	\mbox{$ltl(n,w) \cup subset\{tl(n,p), t<s_{n,w}\}, p<w$} 
\end{array}
\right. \]

The global timeline of a world is therefore the recursive aggregation of the local timeline of the world $w$ with the subset of the global timeline of its parent $p$, up until the divergent point $s_{n,w}$.

Finally, we extend the functions \texttt{insert} and \texttt{read} as:

\medskip
\qquad $insert(c_{n,t,w}, n, t, w)\colon (c \times N \times T \times W) \mapsto Void$, the function that inserts a state chunk in the local timeline of node $n$ and world $w$, such as $ltl_{n,w} := ltl_{n,w} \cup \{c_{n,t,w}\}$.
\medskip

$read(n,t,w)\colon (N \times T \times W) \mapsto c$ is the function that retrieves a state chunk from a world $w$, at time $t$. 
It is recursively defined as:
\small
\[\mbox{$read(n,t,w)$} = \left\{
\begin{array}{lll}
	\mbox{$read_{ltl_{n,w}}(n,t)$} & \mbox{if $(t \geq s) \wedge (ltl_{n,w}\neq \emptyset)$} \\
	         \mbox{$read(n,t, p)$} & \mbox{if ($t<s) \wedge (p <w, p \neq \emptyset) $} \\
	                     \emptyset & \mbox{otherwise}
\end{array}
\right. \]
\normalsize
\medskip

The function \texttt{insert} always operates on the local timeline $ltl_{n,w}$ of the requested node $n$ and world $w$.  
For the function \texttt{read}, if the requested time $t$ is higher or equal to the divergent point in time $s_{n,w}$ of the requested world $w$ and node $n$, the read is resolved on the local timeline $ltl_{n,w}$, as defined in Section~\ref{sec:graph_semantic:temporal_graph_model}. 
Otherwise, we recursively resolve on parent $p$ of $w$, until we reach the corresponding parent to read from. 
 
Once the world resolution is completed, a MWG state chunk can be reduced to a temporal graph state chunk, which in turn can be reduced to a base graph state chunk once the timepoint is resolved.
Similarly, over all nodes, a MWG can be reduced to a temporal graph, then to a base graph, once the read function dynamically resolves the world and time. 

\begin{figure}	
 	\centering			
 	\includegraphics[width=.85\linewidth]{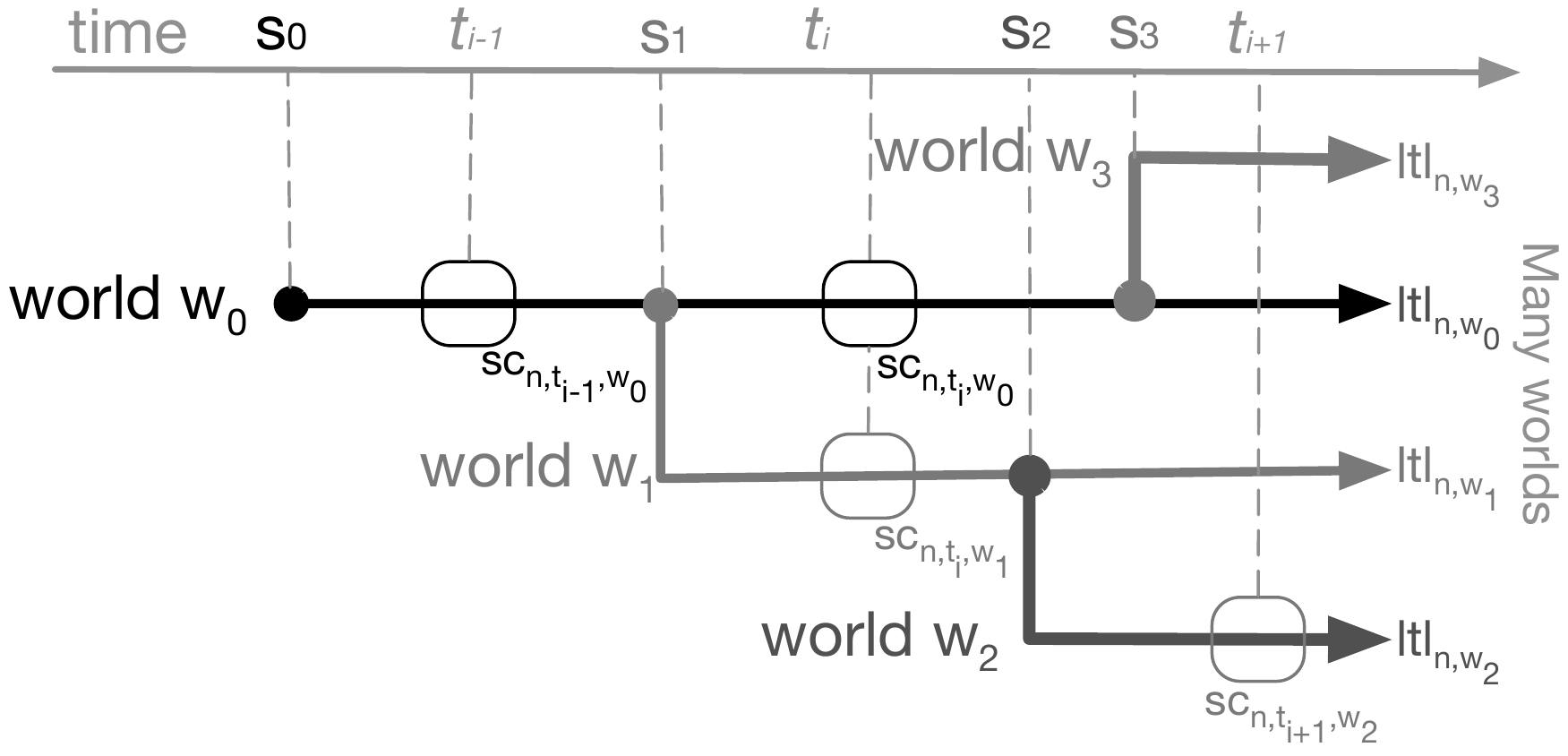}
 	\caption{Many worlds example}
 	\label{fig:parralelworlds}  
 \end{figure}

Figure~\ref{fig:parralelworlds} shows an example of a MWG with several worlds. 
$w_0$ is the root world. In this figure, $w_1$ is diverged from $w_0$, $w_2$ from $w_1$, and $w_3$ from $w_0$. Thus we have the following partial order: $w_0<w_1<w_2$ and $w_0<w_3$.
But no order between $w_3$ and $w_2$ or between $w_3$ and $w_1$. $s_{i}$ for i from 0 to 3, represent the divergent timepoint for world $w_i$ respectively. 
An insert operation on the node $n$ and in any of the worlds $w_i$, will always insert in the local timeline $ltl_{n,w_i}$ of the world $w_i$. 
However, a read operation on the world $w_2$ for instance, according to the shared-past view, will resolve a state chunk from $ltl_{n,w_2}$ if $t\geq s_2$, from $ltl_{n,w_1}$ if $s_1 \leq t < s_2$, from $ltl_{n,w_0}$ if $s_0 \leq t < s_1$, and $\emptyset$ if $t<s_0$.

It is important to note that this semantics goes beyond copy-on-write strategies~\cite{Fabrega95copyon} as a world is never copied, even if data is modified.
Instead, only modified nodes are copied and transparently loaded. 
In fact, this is similar to the concept how we organise and represent the temporal data aspect of the graph.

\section{GreyCat: Implementing MWG}\label{sec:middleware_implementation}
Our MWG concept is supported by an implementation, named \SYS{}, to create, read, update, fork and delete graphs and nodes along time.
In particular, the following sections dive into the implementation details of \SYS{} to expose the design choices we made to outperform the state-of-the-art.

\subsection{Mapping Nodes to State Chunks}\label{sec:graph_serialization}
\begin{figure}
	\centering
	\includegraphics[width=\linewidth]{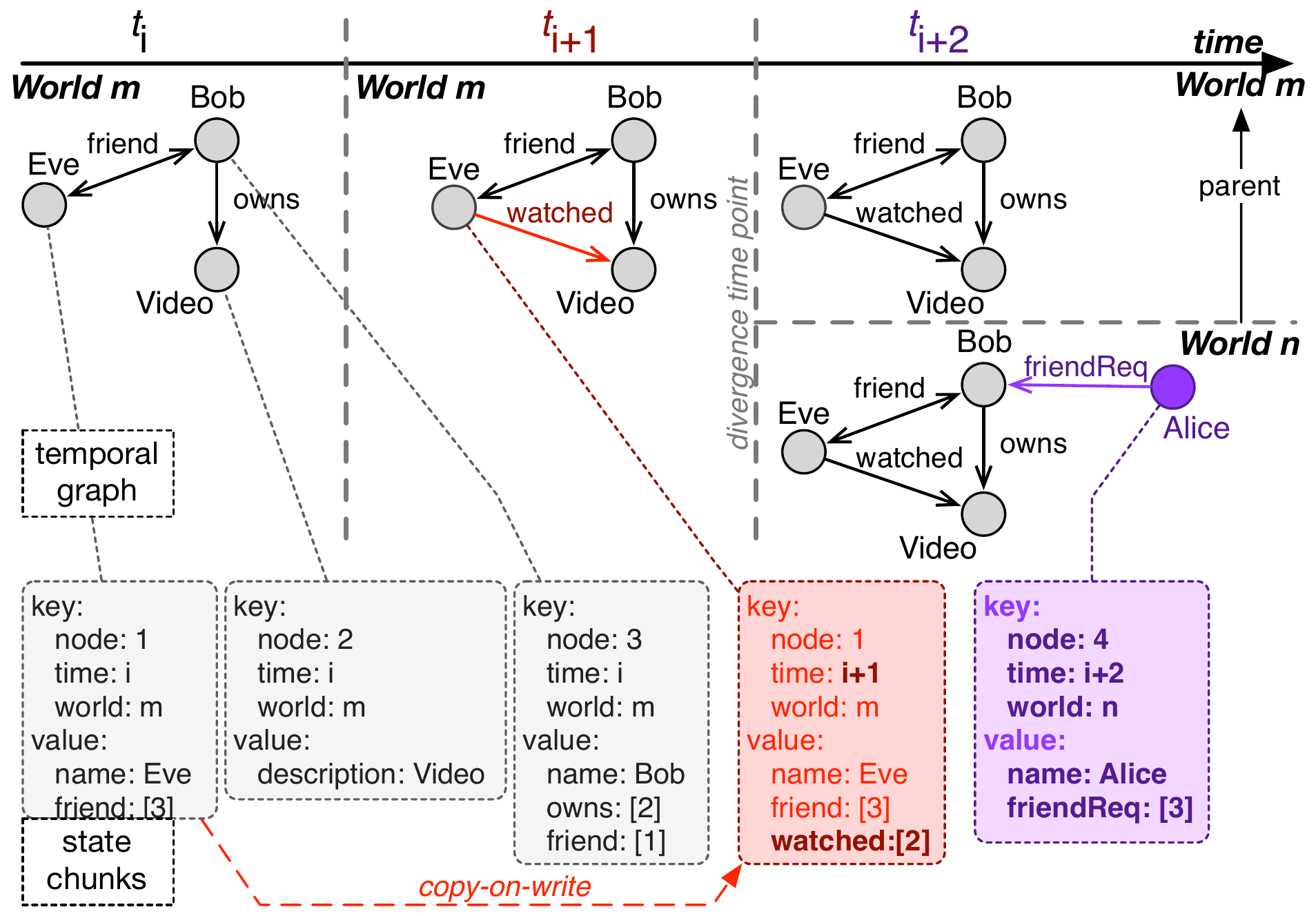}
	\caption{Mapping of nodes to storable state chunks}\label{fig:mapping_nodes_to_statechunks}
\end{figure}

The MWG is a conceptual view of data to work with temporal data and to explore many different alternative worlds.
Internally, we structure the data of a MWG as an unbounded set of \textit{state chunks}.
Therefore, as discussed in Section~\ref{sec:overview}, we map the conceptual nodes (and relationships) of a MWG to \textit{state chunks}.
State chunks are the internal data structures reflecting the MWG and at the same time also used for storing the MWG data.
A state chunk contains, for every attribute of a node, the name and value of the attribute and, for every outgoing relationship, the name of the relationship and a list of identifiers of the referenced state chunks.
Figure~\ref{fig:mapping_nodes_to_statechunks} depicts, in form of a concrete example, how nodes are mapped to state chunks in accordance with the semantic definitions of Section~\ref{sec:manyworld_graph_semantic}.

At time $t_i$ (the starting time of the MWG), \SYS{} maps the nodes and the relationships to 3 state chunks: one for \textsf{Eve}, one for \textsf{Bob}, and one for Bob's \textsf{video}.
At time $t_{i+1}$, the MWG evolves to include a relationship \textsf{watched} from \textsf{Eve} to \textsf{Bob's video}.
Since this evolution only affects \textsf{Eve}, \SYS{} only creates an additional state chunk for \textsf{Eve} from time $t_{i+1}$.
All other nodes are kept unchanged at time $t_{i+1}$ and thus still valid.
Then, at time $t_{i+2}$, world $m$ of the MWG diverges into two worlds: $m$ and $n$.
While world $m$ remains unchanged, in world $n$ \textsf{Bob} meets \textsf{Alice}, who sends a friend request to \textsf{Bob}.
Only \textsf{Alice} changes so that \SYS{} only creates one state chunk for \textsf{Alice} from time $t_{i+2}$ and world $n$.
Here, we see the benefit of the MWG and its semantics: while we are able to represent complex graph topologies, which evolve in time and in many worlds, we only need to store a fraction of this structure.
In this example, the graph contains $13$ different nodes and $16$ relationships (counting each bidirectional relation as two) and evolves along $2$ different worlds and $3$ different timestamps, but we only have to create $5$ state chunks to represent all of this.
Whenever the MWG is traversed, the correct state chunks are retrieved with the right time and world.
The resolution algorithm behind this is presented in Section~\ref{sec:graph_resolution_alg:alg}.

State chunks are the storage units of \SYS{}.
They are stored on disk and loaded into main memory while the MWG is traversed or when nodes are explicitly retrieved.
Loading state chunks can be qualified as lazy, because only attributes and sets of identifiers are loaded.
This theoretically allows to process unbounded MWGs.
For persistent storage of state chunks, we rely on key/value stores by using the 3-tuple of $\{node; time; world\}$ as key and the state chunk as value.
We serialize chunk states into Base64 encoded \texttt{blob}s.
Despite being simple, this format can be used to distribute state chunks over networks.
Moreover, it reduces the minimal required interface to insert state chunks into, and read from, a persistent data store to \texttt{put} and \texttt{get} operations.
This allows to use different storage backends depending on the requirements of an application: from in-memory key/value stores up to distributed and replicated NoSQL databases.


This mapping approach copies state chunks only on-demand---\emph{i.e.}, on-write (per time and world).
This delivers very efficient read and write operations at any point in time.
Using diffs instead of our proposed on-demand forking concept could---in some cases---save disk space, but it would come with a much higher cost for inserting and reading.

\subsection{Indexing and Resolving Chunks}\label{sec:graph_resolution_alg}
This section focuses on the index structures used in \SYS{} and the state chunk resolution algorithm. 
In particular, \SYS{} combines two structures for the indexes of the MWG: \emph{time trees} and \emph{many-world maps}.

\subsubsection{Index Time Tree (ITT)}\label{sec:timetree}
\begin{figure}	
	\centering			
	\includegraphics[width=.7\linewidth]{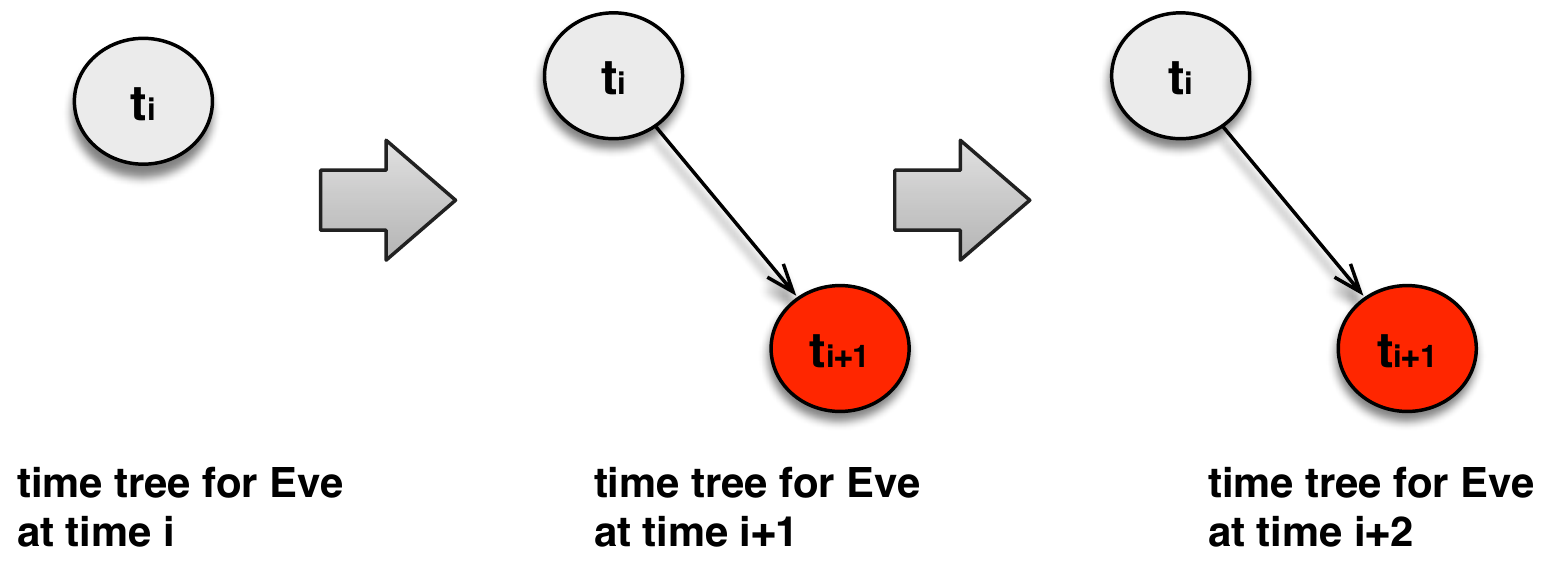}
	\caption{Example ITTs for node \textit{Eve} of Figure~\ref{fig:mapping_nodes_to_statechunks}}
	\label{fig:time_trees}  
\end{figure}

As discussed in Section~\ref{sec:graph_semantic:simple_graph_model}, timepoints are chronologically ordered.
This creates implicit intervals of ``validities'' for nodes in time.
Finding the right ``position'' in a timeline of a node has to be very efficient.
New nodes can be inserted at any time---\emph{i.e.}, not just after the last one.
Besides, ordered trees ({\em e.g.,} binary search trees) are suitable data structures to represent a temporal order, since they have efficient random insert and read complexities. 
If we consider $n$ to be the total number of modifications of a node, the average insert/read complexity is $O(log(n))$ and $O(n)$ in the worst case (inserting new nodes at the end of a timeline).
Given that inserting new nodes at the end of a timeline and reading the latest version of nodes is the common case, we use red-black trees~\cite{Guibas:1978:DFB:1382432.1382565} for the implementation of our time tree index structure.
The self-balancing property of red-black trees avoids the tree  to only grow in depth and improves the worst case of insert/read operations to $O(log(n))$.
Furthermore, we used a custom Java implementation of red-black trees, using primitive arrays as a data backend to minimize garbage collection times, as garbage collection can be a severe bottleneck in graph data stores~\cite{Veiga:2015:ACG:2814576.2814813}. 
Every conceptual node of a MWG can evolve independently in time. 
For scalability reasons, we decided to use one red-black black tree, further called \textit{index time tree} (ITT), per conceptual node to represent its timeline.
Figure~\ref{fig:time_trees} depicts how the ITT looks like and evolves for the node \textit{Eve} introduced in Figure~\ref{fig:mapping_nodes_to_statechunks}.

 \label{subsec:index_time_tree}

As it can be seen, at time $t_i$, one conceptual version of node \textit{Eve} exists and therefore the ITT has only one entry. 
At time $t_{i+1}$, \textit{Eve} changes, a new conceptual version of this node is created and the ITT is updated, accordingly. 
Then, at time $t_{i+2}$, there are additional changes on the MWG, which do not impact \textit{Eve}: the ITT of \textit{Eve} remains unchanged.

ITTs are special state chunks and stored/loaded in the same way to/from key/value stores than any other state chunk. 
More specifically, as key, we use the \textit{id} of the corresponding conceptual node, together with the \textit{world identifier} and the \textit{type} (time tree in this case). 
The value is a serialized and Base64 encoded blob of the tree's values.

\subsubsection{World Index Maps (WIM)}\label{wim}
Since new worlds can diverge from existing worlds at any time and in any number, the hierarchy of worlds can arbitrarily grow both in depth and width. 
As it can be observed in Figure~\ref{fig:parralelworlds}, the divergent point is therefore not enough to identify the parent relationship. 
The divergent point is therefore not enough to identify the parent relationship. 
In our many-world resolution, we use a global hash map that stores, for every world $w$, the corresponding parent world $p$ from which $w$ is derived: $w \rightarrow p$.
We refer to it as the \textit{global world index map} (GWIM).
This allows \SYS{} to insert the parent $p$ of a world $w$, independently of the overall number of worlds, in average in constant time $O(1)$ and in the worst case in $O(l)$, where $l$ is the total number of worlds. 
We also use a custom Java hash map implementation built with primitive arrays to minimize garbage collector effects. 

In addition to the GWIM, \SYS{} defines one local index map, called \textit{local world index map} (LWIM), per conceptual node to identify different versions of the same conceptual node in different worlds.
In this map, we link every world in which the node exists with its ``local'' divergent time, meaning the time when this node was first modified (or created) in this world and therefore starts to diverge from its parent: $w \rightarrow t_{local\ divergence}$.
As we will see, this information is needed to resolve state chunks.
The LWIM is the core allowing nodes to evolve independently in different worlds.
When a conceptual node is first modified (or created) in a world, its state chunk is copied (or created) and the LWIM of the node is updated---{\em i.e.}, the world in which the node was modified is inserted (and mapped to its local divergence time). 
Both the GWIM and the LWIM must be recursively accessed for every read operation of a node (see the semantic definitions in Section~\ref{sec:graph_semantic:manyworld_graph_model}).

\begin{figure}
	\centering			
	\includegraphics[width=\linewidth]{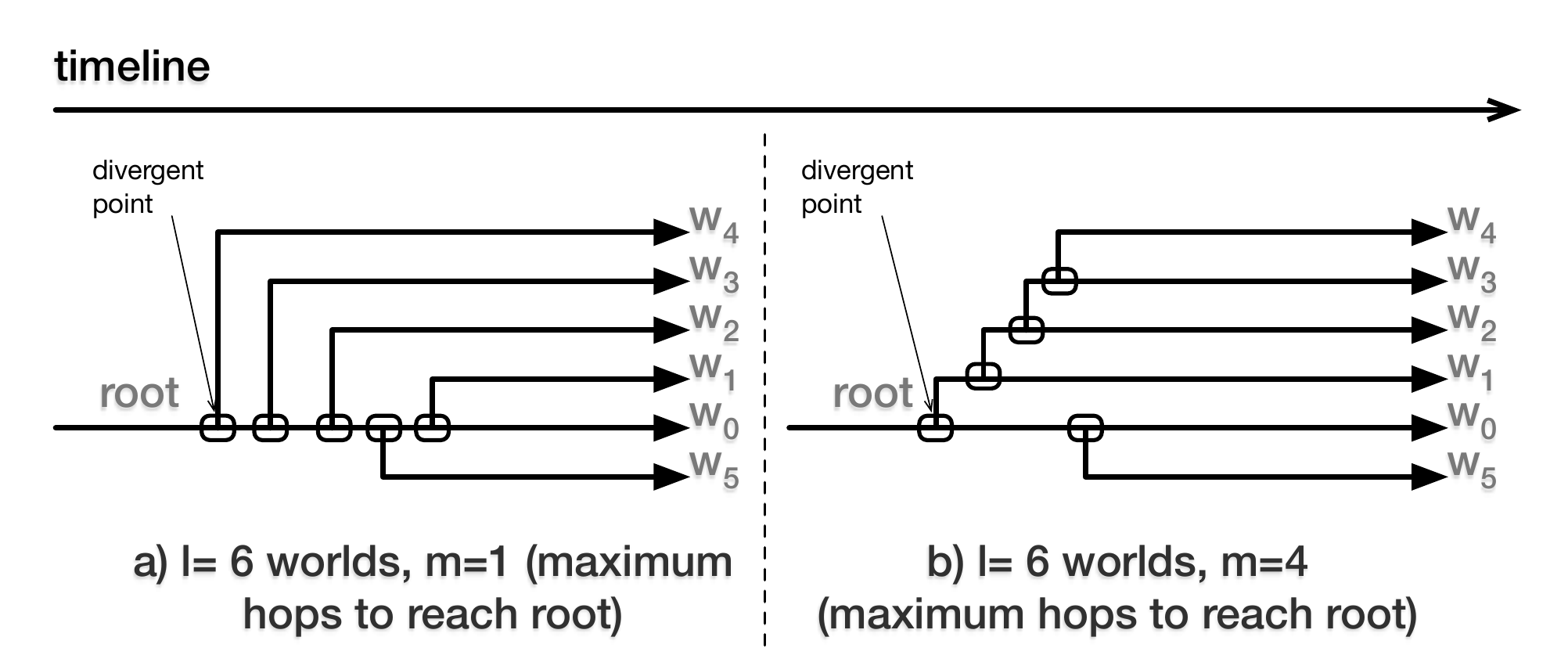}
	\caption{Example of different configurations of the same number of worlds $l$, but with a different $m$}
	\label{fig:hops}  
\end{figure}

Other than the total number of worlds $l$, we define another notation: $m$, as the maximum number of hops necessary to reach the root world ($m \leq l$).
Figure~\ref{fig:hops} reports on an example of 2 MWGs with the same number of worlds $l=6$ but, in the first case we can always reach the root world in $m=1$ hop, while in the second case, we might need $m=4$ hops in the worst case (from world $w_4$ to $w_0$).

The recursive world resolution function has a minimum complexity of $O(1)$ in the best case, where all worlds are directly derived from the root world (shown in Figure~\ref{fig:hops}-a).
The worst case complexity is $O(m) \leq O(l)$, like for the \textbf{stair-shaped} case shown in Figure~\ref{fig:hops}-b, where we might to have to go several hops down before actually resolving the world. 

Like it is the case for ITTs, WIMs are special state chunks and stored/loaded in the same way than regular state chunks.

\subsubsection{Chunk Resolution Algorithm}\label{sec:graph_resolution_alg:alg}
To illustrate the resolution algorithm of MWG, let us consider the example of Figure~\ref{fig:mapping_nodes_to_statechunks}, assuming we want to resolve node \textit{Bob} at time $t_{i+2}$ in world $n$.
We first check the LWIM of \textit{Bob} and see that there is no entry for world $n$, since \textit{Bob} has never been modified in this world.
Therefore, we resolve the parent of world $n$ with the GWIM, which is world $m$.
A glance in the LWIM of \textit{Bob} reveals that world $m$ diverged (or started to exist in this case) for \textit{Bob} at time $t_i$.
This indicates MWG that world $m$ is the correct place to lookup state chunks, since we are interested in \textit{Bob} at time $t_{i+2}$, which is after time $t_i$ where world $m$ for \textit{Bob} becomes valid.
World $m$ is the ``closest'' where \textit{Bob} has been actually modified. 
Otherwise, it would have been necessary to recursively resolve the parent world of $m$ from the GWIM until we find the correct world.
In a last step, we look at the ITT of \textit{Bob} to find the ``closest'' entry to time $t_{i+2}$, which is time $t_i$.
Finally, this index indicates \SYS{} to resolve the state chunk for \textit{Bob} (id 3) with the following parameters: $\{node\ 3; time\ i; world\ m \}$. 
This state chunk resolution is summarized in Algorithm~\ref{alg:state_chunk_resolution}.

 \begin{algorithm}
 \caption{State chunk resolution} \label{alg:state_chunk_resolution}
 \begin{algorithmic}[1]
   \small
 	\Procedure{resolve}{$id, t, w$} 
 		\State $lwim \gets getLWIM(id)$ 
 		\State $s \gets lwim.get(w)$ 
	
 		\If{t >= s} 
 			\State $itt \gets getITT(id)$ 
 			\State \textbf{return} $itt.get(t, w)$ 
 		\Else
 			\State $p \gets GWIM.getParent(w)$ 
 			\State \textbf{return} $resolve(id, t, p)$
 		\EndIf
 \EndProcedure
 \end{algorithmic}
 \end{algorithm}

The full resolution algorithm has a complexity of $O(1)$ for insert, and a complexity of $O(1)+O(m) + O(n) \leq O(l)+O(n)$ for read operations, where $l$ is the number of worlds, and $n$  the number of time points, and $m$ the maximum depth of worlds.

\subsection{Scaling the Processing of Graphs}\label{sec:transactions}
Memory management and transactions or ``units of work'' are closely related.
In \SYS{}, we first connect our graph to a database.
This connection, further called \textit{unit of work} (UoW), marks the beginning of what can be seen in a broader sense as a long-living transaction.
While working with this connection, the state chunks representing the MWG are loaded on-demand into main memory.
All modifications of the MWG are performed in memory.
When saving, the modified and new state chunks (internally marked as \textit{dirty}) are written from memory into (persistent) key/value stores.
Then, the allocated memory is tagged as selectable for eviction, which marks the end of the UoW.
This, together with the on-demand loading of state chunks into main memory, allows to work with graphs of unlimited sizes, in theory.
To increase the read performance, \SYS{} uses local eviction-based LRU caches~\cite{conf/usenix/ChenZL03}.


\subsection{Concurrency and Consistency}\label{sec:concurrency}
In this section, we discuss concurrency and consistency properties of \SYS{}. This section focuses on multi-core architectures while Section~\ref{sec:distribution} describes these properties for distributed deployments.

\textbf{Concurrency}: in order to ensure concurrency, \SYS{} uses a per-node locking strategy for every operation that impacts the timeline of a particular node---\ie new time or world insertion.
For such a major operation, it is important to notice that we lock the whole conceptual node rather than only one state chunk, belonging to one precise world and time.
This means that all data structures---\ie all state chunks and index structures belonging to the locked node, are locked for temporal or many-world write operations.
This way, temporal indexes are always consistent.
Nonetheless, for other write operations, such as attribute value modification or relationship insertion, we adopt a lock per state chunk.
Therefore, parallel writes are allowed if they do not modify the node timeline or the same time-point.
To keep achieving a high level of parallelism, only write operations are blocked while concurrent read operations remain---with some restrictions---allowed.
More specifically, \SYS{} uses a \textit{compare-and-swap} mechanism on the LWIM ensuring that a read operation of a specific node (for a given time and given world) is blocked in case that the requested node is concurrently modified in the same world.
Otherwise, in the case of read operations for the same node in another world, are fully concurrent without locking.

\textbf{Consistency}: the previously defined locking strategy of \SYS{} aims at ensuring consistency per node under parallel read and write operations.
In addition, \SYS{} indexes are created as node attributes, therefore following the locking scheme indexes are consistent by using the standard locks like for write operations.
Based on this decomposition of \SYS{} indexes, no global consistency strategy is necessary.

\subsection{Distribution}\label{sec:distribution}
\lstset{
label=lstClassifier, basicstyle=\footnotesize\ttfamily, language=Java, showstringspaces=false, keywordstyle=\color{blue}, morekeywords={@Override}
}

\SYS{} defines a data access layer that can be distributed over various computers.
However, the current implementation of \SYS{} does not define a specific sharding mechanism to distribute homogeneously data stream on a pool of computers.
Instead, \SYS{} relies on the underlying key/value store for distributed storage of our graph.
Depending on the application requirements, different key/value stores can be plugged via a simple interface, which basically only relies on the implementation of a \texttt{get} and \texttt{put} methods.
For example, if performance is the most critical requirement, but fault tolerance, availability, distribution, and replication are less important, \SYS{} provides drivers for \textsc{RocksDB}~\cite{rocksdb} and \textsc{LevelDB}~\cite{LevelDB}.
On the other hand, if availability, scalability, and replication are more critical, \SYS{} provides also implementations for drivers for \textsc{Cassandra}~\cite{cassandra} and \textsc{HBase}~\cite{george2011hbase}.

To interconnect heterogeneous platforms, such as Android mobile devices, browser environments, and Java-powered servers, we use buses based on WebSocket or MQTT protocols.
This implies a distribution consistency mechanism.
To avoid the use of distributed locks, which could drastically decrease performance, we decided to use an optimistic approach using \textit{Conflict-Free replicated Data Types} (CRDT).
By using CRDT structures for every chunk, we ensure the ability to merge every distributed concurrent modification in a consistent manner.
In addition, we are studying the extension of such mechanism by using a \textsc{Raft} algorithm to offer, on-demand, a consensus primitive to ensure distributed consistency for a dedicated zone of the temporal graph.

\subsection{Working with MWGs}
In order to work with MWGs---\ie to create, navigate, and analyse them---\SYS{} provides Java APIs for developers.
For example, Listing~\ref{lst:creating_graphs} illustrates how a graph can be created using \SYS{}.

\begin{algorithm}[thbp]
	\newalgname{Listing}
	\caption{Java API to create a MWG}
	\label{lst:creating_graphs}
	\begin{lstlisting}[escapechar=!]
public static final long TIME = 0;
public static final long WORLD = 0;

Graph g = new GraphBuilder().build();
g.connect(new Callback<Boolean>() {
  @Override
  public void on(Boolean result) {
    Node eve = graph.newNode(WORLD, TIME);
    node.set("name", Type.STRING, "Eve");
    Node bob = graph.newNode(WORLD, TIME);
		node.set("name", Type.STRING, "Bob");
    eve.addToRelation("friend", bob);
		bob.addToRelation("friend", eve);
  }
});
	\end{lstlisting}
\end{algorithm}

The listing creates a graph consisting of two nodes, \textsf{eve} and \textsf{bob}.
Both are created in world \texttt{0} and for time \texttt{0}.
Node \textsf{eve} has one attribute \texttt{name}, which is of type \texttt{String} and is set to \texttt{Eve}.
In addition, it has a relation named \texttt{friend} to node \texttt{bob}.

Any graph can be manipulated and evolve over time.
Listing~\ref{lst:temporal_modifications} depicts how \SYS{}'s API can be used to do this, by extending the example of Listing~\ref{lst:creating_graphs}.

\begin{algorithm}[thbp]
	\newalgname{Listing}
	\caption{Java API to change a graph over time}
	\label{lst:temporal_modifications}
	\begin{lstlisting}[escapechar=!]
public static final long TIME = 0;

Graph g = new GraphBuilder().build();
g.connect(new Callback<Boolean>() {
  @Override
  public void on(Boolean result) {
    //... code from Listing 2
    long newTime = TIME + 100;
    eve.travelInTime(newTime, node -> {
      node("age", Type.INTEGER, 18);
		});
  }
});
	\end{lstlisting}
\end{algorithm}
The listing moves the node \texttt{eve} to \texttt{TIME + 100} and changes its attribute \texttt{age} to \texttt{18}.
This means, form time \texttt{TIME + 100} on, \texttt{age} will be resolved as \texttt{18}, before this time, it will be resolved to \texttt{17}.

Now, let us consider several different worlds in Listing~\ref{lst:world_modifications}.
We are interested in diverging a world \texttt{1} from world \texttt{0} and change the \texttt{name} of Eve in this divergent world to \texttt{Alice}.

\begin{algorithm}[thbp]
	\newalgname{Listing}
	\caption{Java API to change a graph in several different worlds}
	\label{lst:world_modifications}
	\begin{lstlisting}[escapechar=!]
Graph g = new GraphBuilder().build();
g.connect(new Callback<Boolean>() {
  @Override
  public void on(Boolean result) {
    //... code from Listing 2
    long newWorld = 1;
    eve.travelInWorld(newWorld, node -> {
      node("name", Type.STRING, "Alice");
		});
  }
});
	\end{lstlisting}
\end{algorithm}

In order to process and analyse such a temporal graph, \SYS{} provides an API to efficiently navigate and query the content of the graph.
From any given node, the graph can be easily traversed, as shown in Listing~\ref{lst:graph_traversal}.

 \begin{algorithm}[thbp]
	\newalgname{Listing}
	\caption{Java API to traverse a graph}
	\label{lst:graph_traversal}
	\begin{lstlisting}[escapechar=!]
Graph g = new GraphBuilder().build();
g.connect(new Callback<Boolean>() {
  @Override
  public void on(Boolean result) {
    //... code from Listing 2
    eve.relation("friend", new Callback<Node[]>(){
      @Override
      public void on(Node[] friends) {
        //...
      }
    });
	}
});
	\end{lstlisting}
\end{algorithm}

Since most operations in \SYS{} are non-blocking and therefore asynchronous, deep navigations inside a graph can lead to many nested callbacks (\cf Listing~\ref{lst:graph_traversal}).
Therefore, \SYS{} offers what we call a \textit{Task API}, which is comparable to \texttt{Promises} and \texttt{Futures} and allows the developer to chain several navigation operations without the need to nest them.
In addition, this Task API allows one to specify if a given number of \texttt{Task}s can be executed in parallel and also if they should be executed on a specific machine, \eg local or on a remote machine.
Therefore, tasks, their parameters, and results must be serializable.

While Listing~\ref{lst:graph_traversal} shows how the graph can be navigated from a specific node, the question remains how we find this starting point in the first place.
To solve this issue, \SYS{} uses indexes, which can be created and queried as shown in Listing~\ref{list:graph_query}.

 \begin{algorithm}[thbp]
	\newalgname{Listing}
	\caption{Java API to query a graph}
	\label{list:graph_query}
	\begin{lstlisting}[escapechar=!]
//...
// creating index "nameIndex" from attribute "name"
 g.index(WORLD, Constants.BEGINNING_OF_TIME, "nameIndex", new Callback<NodeIndex>() {
   @Override
   public void on(NodeIndex index) {
     index.addToIndex(self, "name");
   }
 });

// using indexes
indexNode.find(nodes -> {
	// filter here
});
	\end{lstlisting}
\end{algorithm}

As can be seen in the listing, the index itself is a regular node and can also evolve over time.

These examples are showing only the main concepts of \SYS{}'s API's to work with MWG.
More advanced queries and navigation methods are available.
The complete API can be found online on GitHub\footnote{https://github.com/datathings/greycat}.

\section{Experiments}\label{sec:evaluation}
This section reports on the extensive experiments we performed to assess the performance of \SYS{} along several perspectives, from reporting on a real industrial case study (cf. Section~\ref{sec:eval:sgcasestudy}), to comparing it against the closest solutions in the state of the art (cf. Sections~\ref{sec:eval:base}--\ref{sec:eval:time}), and to stressing it against micro-benchmarks (cf. Sections~\ref{sec:simplemwg}--\ref{sec:eval:deepwhatif}).

\subsection{Experimental Setup}\label{sec:eval:exp}
For all these experiments, we report on the throughputs of \textbf{insert} and \textbf{read} operations as key performance indicators.
We executed each experiment $100$ times to assess the reproducibility of our results.
Unless stated otherwise, all the reported results are the average of the $100$ executions.
All the experiments have been executed on the \emph{high performance computer} (HPC) of the University of Luxembourg (Gaia cluster)~\cite{hpc}.
We used a Dell FC430 with 2 Intel Xeon E5-2680 v3 processors, running at 2.5\,GHz clock speed and 128\,GB of RAM.
The experiments were executed with Java version 1.8.0\_73.
All experiments (except the comparison to InfluxDB) have been executed in memory without persisting the results.
The rational behind this is that we want to evaluate our \SYS{} implementation and not the performance of 3rd-party key/value stores, which we use for persisting the data.
For similar reasons, caches have been deactivated for all experiments.
All our experiments are available on GitHub~\cite{mwg_experiments}.

All experiments---for \SYS{}, Neo4j, and InfluxDB---are conducted in a non-clustered way, and are comparable, since the goal of the evaluation is to show the capabilities and limits of \SYS{}.
Data sharding and distribution are out of scope of this paper.
We focus on the evaluation of \emph{create}, \emph{read}, and \emph{update} operations since the contributions of this paper are a MWG data model and the associated storage support rather than processing.
Graph processing algorithms, which are based on these primitives, are therefore considered as out of scope of this paper.

\subsection{Smart Grid Case Study}\label{sec:eval:sgcasestudy}
In this experiment, we evaluate \SYS{} on a real-world smart grid case study, which we introduced in Section~\ref{sec:motivation}.
In particular, we leverage MWG to optimize the electric load in a smart grid.
Therefore, we build profiles for the consumption behavior of customers.
Based on the profiles, we simulate different hypothetical what-if scenarios for different topologies, compute the expected electric load in cables, and derive the one with the most balanced load in all cables.
This allows to anticipate which topology is likely to be the best for the upcoming days.

For this experiment, we use an in-memory configuration, without a backend storage, because we do not need to persist all the different alternatives.
We use the publicly available smart meter data from households in London~\cite{london_data_set}.
As the dataset from our industrial partner \textsc{Creos} is confidential, we use this publicly available dataset for the sake of reproducibility.
The grid topology used in our experiments is based on the characteristics of the \textsc{Creos} smart grid deployment~\cite{DBLP:conf/smartgridcomm/0001FKTPTR14}.
We consider $5,000$ households connected to the smart grid, including $4,000$ consumption reports per customer.
This leads to a large-scale graph with $20,000,000$ conceptual nodes used to learn the consumption profiles.
As described in~\cite{DBLP:conf/smartgridcomm/0001FKTPTR14} around $100$ customers are connected to one transformer substation.
We simulate $50$ power substations for our experiments and we suppose that every household can be connected to every power substation.
This is a simplification of the problem, since which household can be connected to which power substation depends on the underlaying physical properties of the grid, which we neglect in the following experiment.

 \begin{figure}
 	\centering
 	\includegraphics[width=\linewidth]{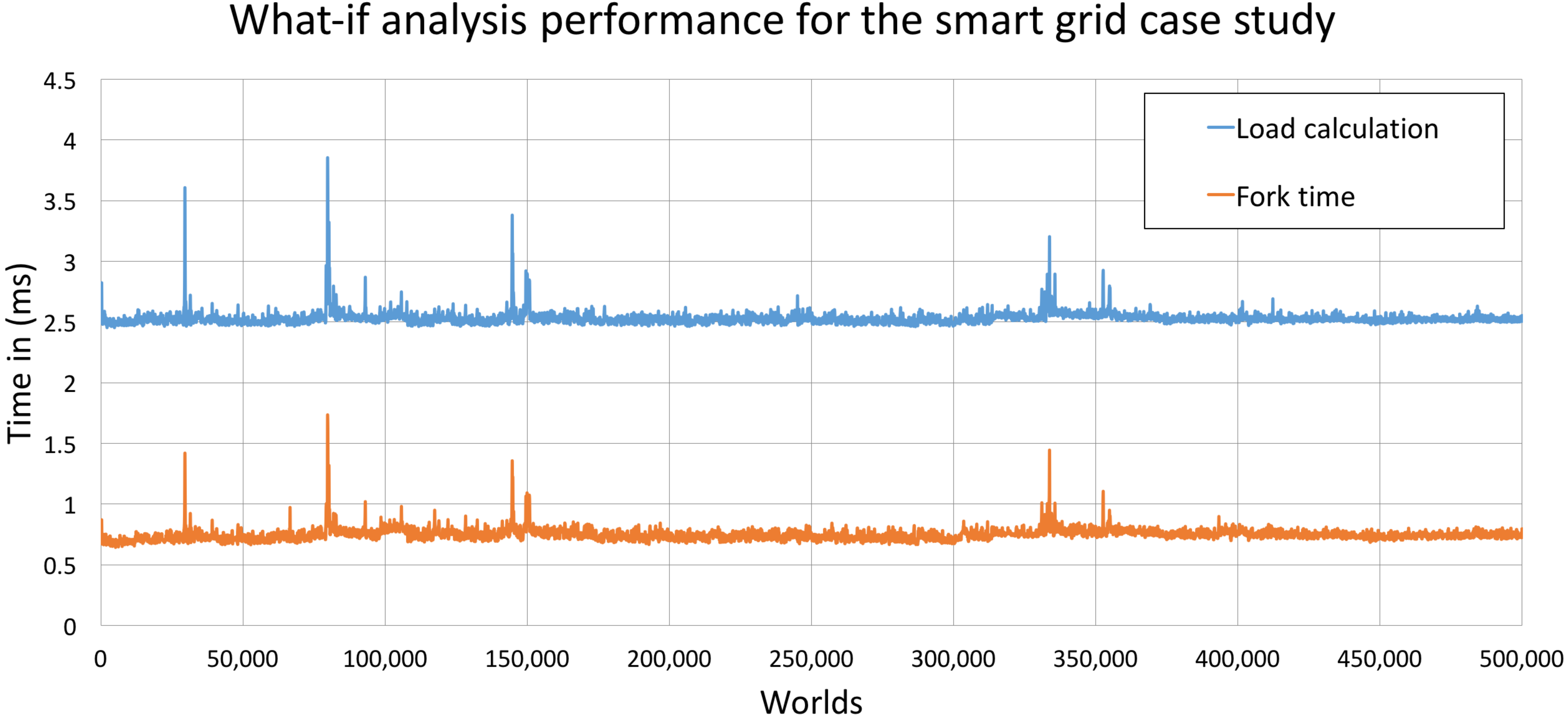}
 	\caption{Performance of load calculation in a what-if scenario for the smart grid case study}
 	\label{fig:finalread}
 \end{figure}

Figure~\ref{fig:finalread} reports on the what-if analysis performed over $500,000$ worlds where, in each world, we mutate 3\,\% of the power substations connections to smart meters.
We plot the latency (in $ms$) of the load calculations and world creation (fork time) per world.
As depicted in Figure~\ref{fig:finalread}, both curves are quite constant, with some peaks due to garbage collection.
Based on this experiment, we can conclude that \SYS{} is scalable and can apply to large-scale systems, such as smart grids.

\subsection{Base Graph Benchmarks}\label{sec:eval:base}
The objective of this benchmark is to evaluate the performance of \SYS{} as a standard graph storage by neglecting time and many-worlds.
Therefore, this section compares the performance of \SYS{} to state-of-the-art graph databases.
For this comparison, we use the graph database benchmark~\cite{graphdb_bench} provided by Beis {\em et al.}~\cite{Beis2015}.
This benchmark is based on the problem of community detection in online social networks.
It uses the public datasets provided by \emph{Stanford Large Network Dataset Collection}~\cite{stanford_dataset}.
This dataset collection contains sets from ``social network and ground-truth communities''~\cite{DBLP:journals/corr/abs-1205-6233}, which are samples extracted from Enron, Amazon, YouTube, and LiveJournal.
The benchmark suite defines several metrics, among which:

\begin{compactdesc}
	\item[\textbf{Massive Insertion Workload} (MIW)] creates the graph database for massive loading, then populates it with a dataset. The creation throughput of the whole graph is reported.
	\item[\textbf{Single Insertion Workload} (SIW)] creates the graph database and loads it with a dataset. Every insertion (node or edge) is committed directly and the graph is constructed incrementally. The insertion throughput is reported.
\end{compactdesc}

 \begin{figure}
 	\centering
 	\includegraphics[width=\linewidth]{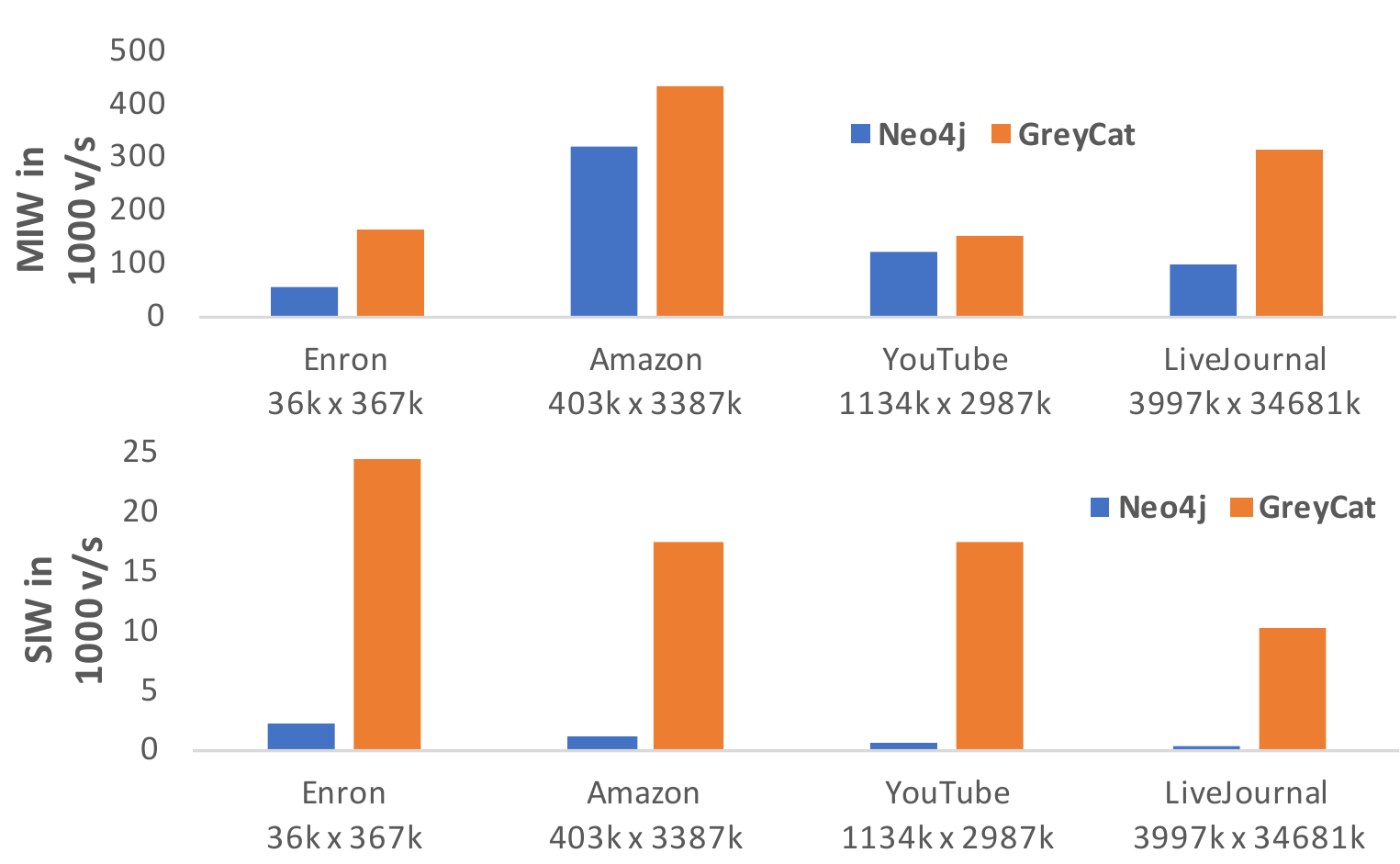}
 	\caption{MIW and SIW benchmark throughput in $1,000$ values/second. Bigger numbers mean better results. Numbers beneath dataset names mean number of nodes and  edges, \emph{i.e. nodes x edges}.}
 	\label{fig:miw_siw_bench}
 \end{figure}

We compare the performance of \SYS{} to Neo4J~\cite{neo4j}, which was the best performing base graph in~\cite{Beis2015}.
Figure~\ref{fig:miw_siw_bench} reports on the results of MIW and SIW, achieved by \SYS{} and Neo4J (both in-memory and not persisting data), along the different datasets.
For all benchmarks, \SYS{} outperforms Neo4J by factors ranging from 1.3x to 20x.

\subsection{Temporal Graph Benchmarks}\label{sec:eval:time}
This experiment aims to evaluate the complexity of the ITT (cf. Section~\ref{sec:timetree}).
We compare the performance of temporal data management of our approach with plain time series databases.
Therefore, we consider only \textbf{one world} and \textbf{one node id} and we benchmark the throughput of \emph{insert} and \emph{read} operations over a varying size of timepoints, from 1 million to 256 million.
Table~\ref{table:ts1} reports on the measured results under progressive load, to check the complexity according to the expected one.

\begin{table}
	\caption{Average \emph{insert} and \emph{read} time, for different timepoints, for one node and in the same world.}
	\label{table:ts1}
	\begin{center}
		\scriptsize
		\begin{tabular}{ l | c c | c c}
			\Xcline{1-5}{0.7pt}
			\textbf{(n) in}& 	\textbf{Insert speed} & \textbf{Read speed} &\textbf{Insert /} & \textbf{Read /}\\
			\textbf{millions}& 	\textbf{($1,000$ val./s)} & \textbf{($1,000$ val./s)} & \textbf{log(n)} & \textbf{log(n)}\\
			\Xcline{1-5}{0.7pt}
			1   & 	589.17 &	605.30 &	42.6	&	43.8	\\
			\Xcline{1-5}{0.5pt}
			2   & 	565.05 &	564.11 &	38.9	&	38.8	\\
			\Xcline{1-5}{0.5pt}
			4   & 	554.40 &	544.23 &	36.4	&	35.8	\\
			\Xcline{1-5}{0.5pt}
			8   & 	537.22 &	528.18 &	33.8	&	33.2	\\
			\Xcline{1-5}{0.5pt}
			16  & 	520.98 &	516.26 &	33.2	&	31.1	\\
			\Xcline{1-5}{0.5pt}
			32  & 	515.05 &	485.73 &	29.8	&	28.1	\\
			\Xcline{1-5}{0.5pt}
			64  & 	489.55 &	458.32 & 	27.2	&	25.5	\\
			\Xcline{1-5}{0.5pt}
			128 & 	423.53 &	400.49 & 	22.7	&	21.5	\\
			\Xcline{1-5}{0.5pt}
			256 & 	391.56 & 	378.50 &	20.2 	&	19.5	\\
			\Xcline{1-5}{0.7pt}
		\end{tabular}
	\end{center}
	\end{table}

As one can observe, \emph{read} and \emph{insert} performance follows an $O(log(n))$ scale as $n$ increases from 1 million to 256 million.
The performance deterioration beyond 32 million can be explained due to a 31 bit limitation in the hash function of the ITT.
This comes from the fact that our ITT is implemented as a red-black tree backed by primitive Java arrays.
These are limited to 31 bit indexes.
At these large numbers, collisions become very recurrent.
For instance, for the 256 million case, there are around 8\,\% of collisions.
This compares to less than $0.02\,\%$ of collisions for 1 million.
To address this problem, we plan for future work an off-heap memory management implementation (based on Java's unsafe operations), which would allow us to solve the limitation of 31 bit indexes for primitive arrays and to use hash functions with more than 31 bits.

To compare with a time series database, namely InfluxDB, we use the influxDB benchmark~\cite{influx_bench}.
It consists of creating $1,000$ nodes (time series) where they insert $1,000$ values in each node on MacBook Pro, resulting in a graph with conceptually $1,000,000$ million nodes.
The second test is to create $250,000$ nodes where they insert $1,000$ values in each, on an Amazon EC2 \textsf{i2.xlarge} instance.
This results in a large-scale graph with conceptually $250,000,000$ nodes.
For both experiments, data is persisted to disk.

The main difference with the experiment above is that the ITT of each node does not grow the same way in terms of complexity as an ITT of 250 million elements in a single node does.
For the sake of comparison, we applied the same benchmarks using the same types of machines.
We use RocksDB~\cite{rocksdb} as our key/value backend.
Despite the fact that our MWG is not limited to flat time series, but a full temporal graph, we are able to outperform InfluxDB by completing the MacBook test in $388$ seconds compared to their $428$ seconds (10\,\% faster), and by getting an average speed of $583,000$ values per second on the amazon instance, compared to their $500,000$ values per second (16\,\% faster).
We note that, when all elements are inserted in the same ITT, the speed drops to $391,560$ inserts per second on average (cf. Table~\ref{table:ts1}).
This is due to the increased complexity of balancing the ITT of one node.
The experiment therefore assess that \SYS{} is able to manage full temporal graphs as efficiently (on a comparable scale) as time series databases are able to manage flat sequences of timestamped values.

\begin{figure}
 	\centering
 	\includegraphics[width=\linewidth]{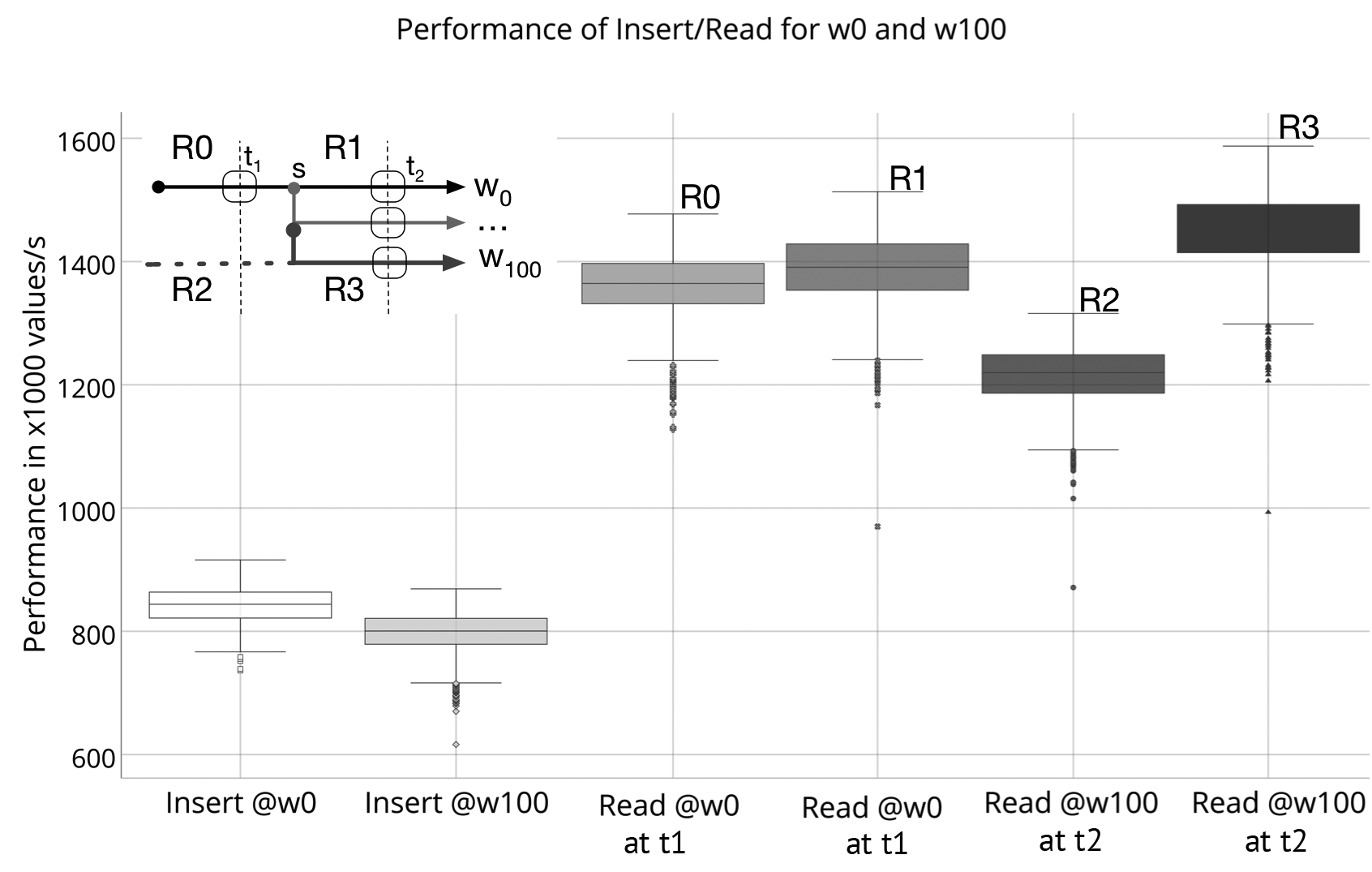}
 	\caption{\emph{Insert} and \emph{read} performance before and after the divergence timepoint \textbf{s}}
 	\label{fig:insertread}
\end{figure}

\subsection{Node-scale Benchmarks}\label{sec:simplemwg}
In this experiment, we demonstrate the effect on \emph{insert} and \emph{read} performance of creating many worlds from one node.
Diverging only one world from the root world is not enough to measure a noticeable performance difference.
Therefore, we created $100$ nested parallel worlds from root world $w_0$.
We first measure the \emph{insert} performance for the worlds $w_0$ and $w_{100}$.
Then, we measure---for the root world---the \emph{read} performance $R_0$ at a shared past timepoint $t_1=5000<s$ and $R_1$ at timepoint $t_2=15000>s$ (after the divergence).
We repeat the experiments for the same timepoints $t_1$ and $t_2$, but from the perspective of world $w_{100}$, to get \emph{read} performance $R_2$ and $R_3$.
The results are depicted in Figure~\ref{fig:insertread} as box plots over $100$ executions.
We can conclude that the \emph{insert} performance is similar for both worlds.
The \emph{read} performance for the root world is not affected by the divergence $R_0=R_1$, while the \emph{read} performance of world $w_{100}$ depends on the timepoint---\emph{i.e.}, it is faster to read after the divergence point than before it ($R_3>R_2$).
This is due to the recursive resolution algorithm of \SYS{}, as explained in Section~\ref{wim}.

In this experiment, we validated that the write and \emph{read} performance on the \SYS{} are not affected by the creation of several worlds.
In particular, we showed that the \emph{read} speed is kept intact after the divergence for the child worlds.

\subsection{Graph-scale Benchmarks}\label{sec:nestedmwg}
\begin{figure}
 	\centering
 	\includegraphics[width=\linewidth]{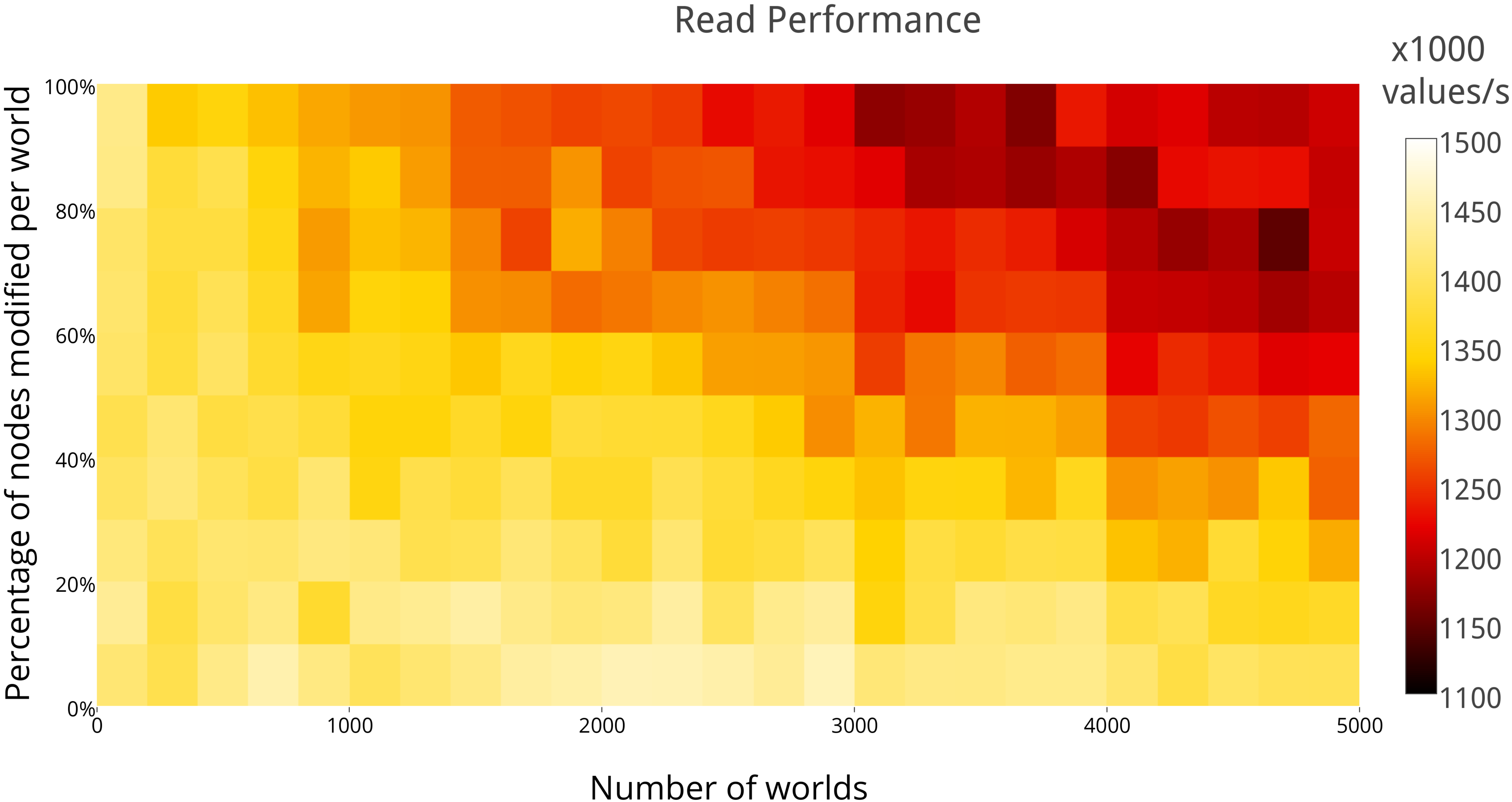}
 	\caption{Read performance, over several worlds and several percentage of modified nodes}
 	\label{fig:nestedread}
 \end{figure}

To stress the effect of recursive world resolution, we consider the \textbf{stair-shaped} scenario presented in Figure~\ref{fig:hops}-b.
In this benchmark, we create a graph of $n=2000$ nodes, each having an initial fixed timeline of $10,000$ timepoints in the main world.
This results in an initial graph of $20,000,000$ conceptual nodes.
Then, we select a fixed $x\,\%$ amount of these nodes to go through the process of creating the shape of stairs of $m$ steps across $m$ worlds.
In each step, we modify one timepoint in the corresponding world of the corresponding node.
For this experiment, we vary $m$ from $1$ to $5,000$ worlds by steps of $200$ and $x$ from 0 to $100\,\%$ per steps of $10\,\%$.
This generates $250$ different experiments.
We executed each experiment $100$ times and averaged the \emph{read} performance of the whole graph before the divergence point, from the perspective of the last world.
Figure~\ref{fig:nestedread} depicts the results as a heat map of the average \emph{read} performance for the different combinations of number of worlds and percentage of nodes changed.
The brightest area (lower left) represent the best performance (low number of worlds or low percentage of nodes changed in each world).
The darkest area (upper right) represent up to $26\,\%$ of performance drop (when facing an high percentage of changes and an high number of worlds).

This benchmark is the worst case for the MWG, since for $m^{th}$ worlds, a \emph{read} operation might potentially require $m$ hops on the WIM, before actually resolving the correct state ({\em e.g.,} reading the first inserted node from the perspective of the last created world), as discussed in Section~\ref{wim}.
The performance drop is linear in $O(m)$ and according to the percentage of nodes changed from one world to another.
For less than 20\,\% of changes, the performance drop is hardly noticeable even at an high number of worlds (lower right).
We note that our solution only stores the modifications for the different worlds and rely on the resolution algorithm to infer the past from the previous worlds.
Any snapshotting technique, cloning the whole graph of $2,000$ nodes, each including $10,000$ timepoints---\emph{i.e.}, a graph with $20,000,000$ nodes---$5,000$ times would be extremely costly to process.
To sum up, we show in this section that our index structure allows independent evolution of nodes at scale.
The performance decreases linearly with the percentage of nodes changed and the maximum of worlds reached.

\subsection{Deep What-if Simulations}\label{sec:eval:deepwhatif}
\begin{figure}
 	\centering
 	\includegraphics[width=\linewidth]{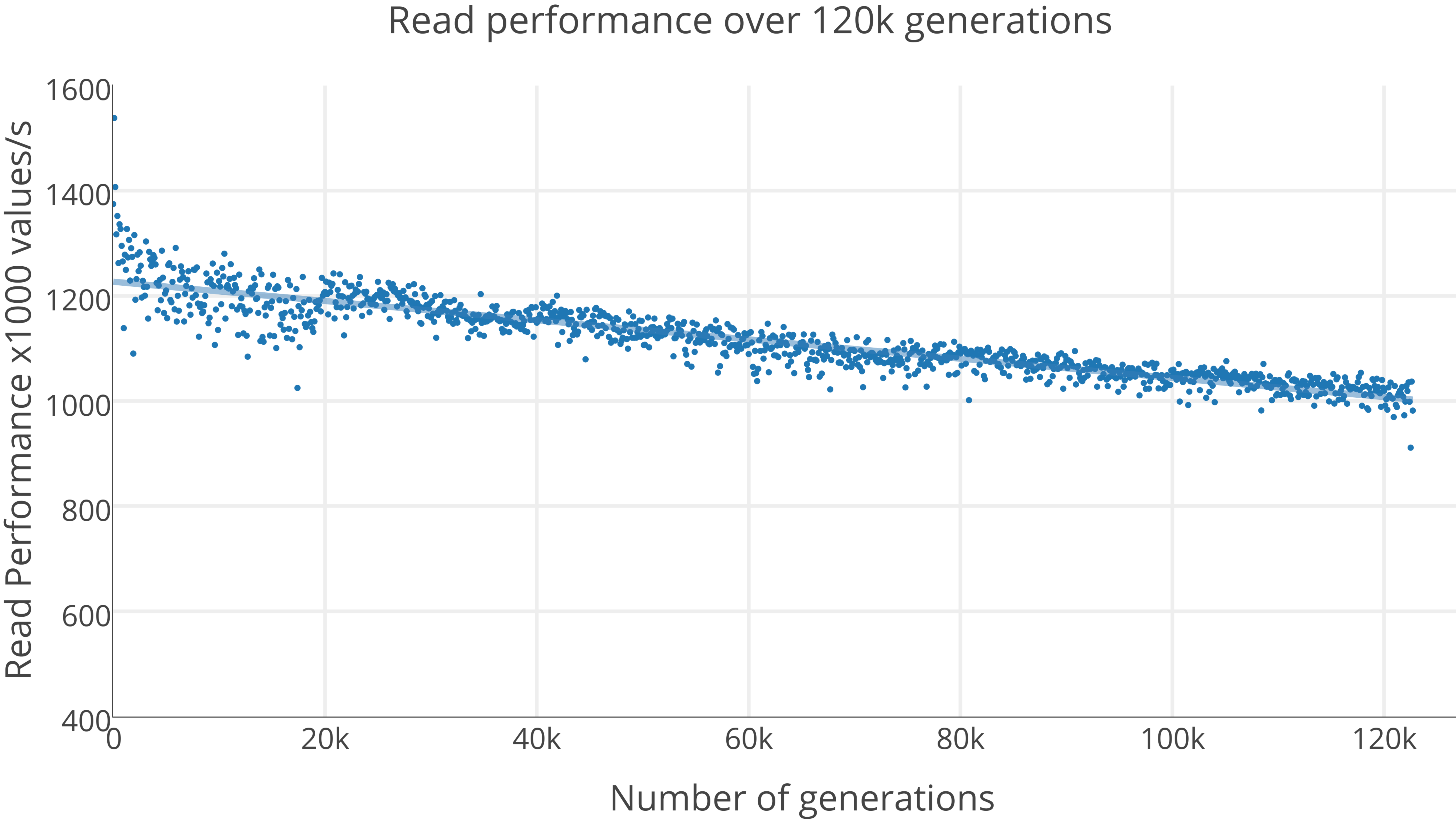}
 	\caption{Average \emph{read} performance over $120,000$ generations with 3\,\% mutations}
 	\label{fig:genetic}
 \end{figure}

As the motivation of our work is to enable deep what-if simulations, we benchmark in this section the \emph{read} performance over a use-case similar to the ones we can find in this domain.
We use a setup similar to the previous section: a graph of $n=2,000$ nodes with initially $10,000$ timepoints in the root world.
The difference is that we fixed the percentage of changes between one world to another to $x=3\,\%$ (similar to a nominal mutation rate in genetic algorithms of 0.1--5\,\%~\cite{srinivas1994genetic}).
The second difference is that changes only randomly affect $3\,\%$ of the nodes for each step.
This is unlike the previous experiment, where the target was to reach a maximum depth of worlds for the same amount of $x\,\%$ of nodes.
We executed this simulation in steps of $1,000$ to $120,000$ generations ($120$ experiments, each repeated $100$ times).
The number of generations is similar to the ones in genetic algorithms~\cite{srinivas1994genetic}.
In each generation, we create a new world from the previous one and randomly modify $3\,\%$ of the nodes.
At the end of each experiment, we measure the performance of reading the whole graph of $1,000$ nodes.
Figure~\ref{fig:genetic} reports on the results MWG achieves.
In particular, one can observe that the \emph{read} performance drops linearly, $28\,\%$ after $120,000$ generations.
This validates the linear complexity of the world resolution, as presented in Section~\ref{wim}, and the usefulness of our approach for what-if simulation when a small percentage of nodes change, even in a huge amount of deep nested worlds.

\label{sec:deepmwg}

\section{Related Work}\label{sec:related}
Over the years, data management systems have pushed the limits of data analytics for huge amounts of data further and further.
In the 1990's Codd {\em et al.}~\cite{codd1993} presented a category of database processing, called \textit{online analytical processing} (OLAP).
It addressed the lack of traditional database processing to consolidate, view, and analyze data according to multiple dimensions.
In a more recent work about best practices for big data analytics, Cohen {\em et al.}~\cite{cohen09madskills} present what they call ``MAD Skills''.
They highlight the practice of magnetic, agile, and deep (MAD) data analysis as a departure from traditional data warehousing and business intelligence.
For example, the Hadoop stack~\cite{hadoop}, Heron~\cite{Kulkarni:2015:THS:2723372.2742788}, and Spark~\cite{zaharia2012resilient} drove the development of new and powerful data analytics.
Despite this, today data analytics is still predominantly \textit{descriptive}. 
However, as Haas {\em et al.}~\cite{HaasVLDB11} suggested, what enterprises really need is \textit{prescriptive} analytics to identify optimal business decisions.
They argue that this requires what-if analysis. 
In accordance with this idea, we propose a graph data model, which is able to efficiently evolve in time and in many worlds to simulate different decisions.

Recently, much work focuses on large-scale graph representation, storage, and processing for analytics.   
Well-known examples are Pregel~\cite{Malewicz:2010:PSL:1807167.1807184}, Giraph, Neo4j~\cite{neo4j}, GraPS~\cite{Canas:2015:GGP:2814576.2814812}, SNAPLE~\cite{Kermarrec:2015:SOL:2814576.2814810}, and GraphLab~\cite{low2014graphlab}.  
While many of them require the graph to be completely in-memory while processing~\cite{ching2015one}, others, like Roy {\em et al.}~\cite{Roy:2015:CSG:2815400.2815408} or Shao \emph{et al.,}~\cite{Shao:2013:TDG:2463676.2467799}, suggest to process graphs from secondary storage. 
Similarly, we allow to store graph data on secondary storage, since even with big clusters at a certain point the limit of in-memory only solutions is reached.
Given that our MWG is built to evolve extensively in time and many worlds, the need for secondary storage is even more underlined, since many different versions of nodes can coexist, making graphs even bigger. 
While most of this work use a rather standard graph data model and focus on graph computation and processing, the focus of our work is rather to support large-scale what-if analysis. 

The need to represent and store the temporal dimension of data has been comprehensively discussed in the database community in the 80s and 90s.  
 For example, Clifford {\em et al.,}~\cite{clifford1983formal} and also Ariav~\cite{ariav1986temporally} provide a formal semantic for historical databases.
In a similar direction goes the work of Ariav~\cite{ariav1986temporally}.
They all suggest, in some way or another, to directly integrate temporal structures in the data model itself, rather than at the application level.
In~\cite{segev1987logical}, Segev and Shoshani discuss the semantics of temporal data and corresponding operators independently from a specific data model.  
Salzberg {\em et al.,}~\cite{Salzberg:1999:CAM:319806.319816} discuss different temporal indexing techniques.
Although most of this work is relatively old, such temporal databases are not very widespread.
Google~\cite{chang2008bigtable} embeds versioning at the core of its BigTable implementation by enabling each cell in a table to contain multiple versions of the same data.

With the emergence of cyber-physical systems, temporal aspects of data evolved again in form of time series management.
As mentioned before, InfluxDB is one of the newer time series databases, which received much attention lately.
They position themselves as an IoT and sensor database for real-time analytics. 
While it provides many interesting features, like a SQL-like query language, their data model is essentially flat and does not support complex relationships between data, \emph{i.e.,} it provides very little support for richer data models, like graphs.
The same counts for Atlas~\cite{atlas}, which was developed by Netflix to manage dimensional time series data for near real-time operational insights, and OpenTSDB~\cite{open_tbsd}.
RRDtool~\cite{rdd_tool} is another data logging and graphing system for time series.
The same counts for Atlas~\cite{atlas}, OpenTSDB~\cite{open_tbsd}, and RRDtool~\cite{rdd_tool}.
All of this work has in common that it provides high performance storage and management specialized for time series data.
However, these solutions provide very little support for richer data models, like graphs.

Lately, an increasing amount of work deals with the need of temporal aspects of graph data. 
Finally, an increasing interest in time-evolving graphs appeared.   
For example, Huanhuan {\em et al.,}~\cite{wu2014path} discuss the problem of finding the shortest path in a temporal graph.
Bahmani {\em et al.,}~\cite{Bahmani:2012:PEG:2339530.2339539} show how to compute PageRank on evolving graphs.
Khurana and Deshpande~\cite{khurana2015storing} present with \emph{Historical Graph Store} (HGS) a system for managing and analyzing large historical traces of graphs.  
T-SPARQL~\cite{grandi2010t} is a temporal extension for the SPARQL~\cite{perez2009semantics} RDF query language. 
HGS consists of two major components, the \emph{Temporal Graph Index} (TGI) and a \emph{Temporal Graph Analysis Framework} (TAF).
Their proposed TGI stores the complete history of a graph in form of partitioned deltas and rebuilds the graph from these deltas while querying graph data.
With TAF, they provide a library to specify a wide range of temporal graph analysis tasks.
With the index time tree, we pursue similar goals as they do with their indexing strategies, however we do not save deltas for the whole graph, but instead treat all versions of nodes similar.
This simplifies the retrieval of historical data without the need to rebuild it from stored deltas.
GraphTau~\cite{Iyer:2016:TGP:2960414.2960419}, Kineograph~\cite{Cheng:2012:KTP:2168836.2168846}, and Chronos~\cite{Han:2014:CGE:2592798.2592799} also extend graph processing to time-evolving graphs.  
While Neo4j itself does not provide any support for temporal data, Cattuto {\em et al.,}~\cite{Cattuto:2013:TSN:2484425.2484442} present a pattern on how to use Neo4j for analyzing time-varying social networks.
They suggest to associate nodes and edges with time intervals (frames) and to represent both logical graph nodes and edges as Neo4j nodes.
This lack of a native support leads to a rather complicated data and query model.
Chronos~\cite{Han:2014:CGE:2592798.2592799} is another interesting storage and execution engine, however it is designed specifically for in-memory iterative graph computation. 
These approaches have in common that they represent time-evolving graphs, in some form or another, as a sequence of snapshots and use a rather standard graph data model.
In addition, most of these approaches requires to keep a full graph snapshot in memory and they have some limits when data is changing at a very high pace.
Our approach suggests an efficient temporal graph data model for large-scale what-if analysis on rapidly changing data.

The idea of what-if analysis with hypothetical queries has been discussed in database communities.
Balmin {\em et al.,}~\cite{Balmin:2000:HQO:645926.672016} proposed an approach for hypothetical queries in OLAP environments. 
They enable data analysts to formulate possible business scenarios, which then can be consequently explored by querying. 
Unlike other approaches, they use a ``no-actual-update'' policy, {\em i.e.,} the possible business scenarios are never actually materialized but only kept in main memory.
In a similar approach, Griffin and Hull~\cite{Griffin:1997:FIH:253260.253304} focus on queries with form \textit{Q when \{U\}} where Q is a relational algebra query.
This paper develops a framework for evaluating hypothetical queries using a ``lazy'' approach, or using a hybrid of eager and lazy approaches.
They present an equational theory and family of rewriting rules that is analogous and compatible with the equational theory and rewriting rules used for optimizing relational algebra queries. 
In~\cite{arenas1998hypothetical} and~\cite{hypo_temporal_reasoning}, Arenas {\em et al.,} developed an approach for hypothetical temporal queries of form \textit{``Has it always been the case that the database has satisfied a given condition C''}. 
Despite there is no explicit time in these queries, they call them ``temporal'' due to a similarity with dynamic integrity constraints.
Although these approaches have a similar goal than our approach, they differ in many major points.
First, they mainly aim at data analysts which perform selective queries on a modest number of possible business scenarios to investigate impacts of decisions. 
In contrary, we aim at intelligent systems and complex data analytics, which need to explore a very large number of parallel actions ({\em e.g.}, as for genetic algorithms or the presented smart grid case study), which can be highly nested. 
Moreover, these systems usually face significantly higher demands regarding performance. 
In addition, most of these approaches do not support (or only in a limited manner) the co-evolution of worlds, which is an essential feature of the MWG.
To the best of our knowledge there is no approach allowing graphs to evolve in time and in many worlds for efficient what-if analysis.
Another major difference is that the MWG is a fully temporal graph supporting both the exploration of different hypothetical worlds and the temporal evolution of data. 
Our proposed MWG can be used in arbitrary analytics and is independent of the concrete underlying database whereas most of the work on hypothetical queries has been done on relational databases.

\section{Conclusion}\label{sec:conclusion}
We proposed a novel graph data model, called \emph{Many-World Graph} (MWG), which allows to efficiently explore a large number of independent actions---both in time and many worlds---on a massive amount of data. 
We validated that \SYS{}, our MWG implementation, follows the theoretical time complexity of $O(log(n))$ for the temporal resolution and $O(m)$ for the world resolution, where $m$ is the maximum number of nested worlds.

Our experimental evaluation showed that even when used as a base graph---without time and many-worlds---\SYS{} outperforms a state-of-the-art graph database, Neo4J, for both mass and single inserts.
A direct comparison with a state-of-the-art time series database, InfluxDB, showed that although the MWG is not just a simple time series, but a fully temporal graph, the temporal resolution performance of MWG is comparable or in some cases even faster than time series databases.
The experimental validation showed that the MWG is very well suited for what-if analysis.
Regarding the support for prescriptive analytics, we showed that \SYS{} is able to handle efficiently hundreds of millions of nodes, timepoints, and hundreds of thousands of independent worlds.

Beyond the specific case of smart grids, we believe that \SYS{} can find applications in a large diversity of application domains, including social networks~\cite{DBLP:conf/middleware/KermarrecTT15}, smart cities, and biology~\cite{HaasVLDB11}.

Aside of potential applications of this approach, our perspectives also include the extension of \SYS{} to consider different \emph{laws of evolution} for the stored graphs, thus going beyond the application of machine learning~\cite{DBLP:conf/smartgridcomm/0001MFRMKT15}.
We also look at the integration of \SYS{} with existing graph processing systems, like Giraph~\cite{Malewicz:2010:PSL:1807167.1807184}.
Finally, beyond the what-if analysis, the coverage of alternative prescriptive analytics based on \SYS{} is a direction we are aiming for.

\section*{Acknowledgments}
The research leading to this publication is supported by the National Research Fund Luxembourg (grant 6816126) and Creos Luxembourg S.A. under the SnT-Creos partnership program.

\section*{Availability}
The source code of \SYS{} is available under:
\begin{center}
{\tt https://github.com/datathings/greycat}\\
\end{center}

\section*{Bibliography}
\bibliographystyle{elsarticle-num}
\bibliography{bibliography}

\begin{thebibliography}{10}
\expandafter\ifx\csname url\endcsname\relax
  \def\url#1{\texttt{#1}}\fi
\expandafter\ifx\csname urlprefix\endcsname\relax\def\urlprefix{URL }\fi
\expandafter\ifx\csname href\endcsname\relax
  \def\href#1#2{#2} \def\path#1{#1}\fi

\bibitem{Kathiravelu:2015:CMP:2836127.2836132}
P.~Kathiravelu, L.~Sharifi, L.~Veiga,
  \href{http://doi.acm.org.proxy.bnl.lu/10.1145/2836127.2836132}{{Cassowary:
  Middleware Platform for Context-Aware Smart Buildings with Software-Defined
  Sensor Networks}}, in: Proceedings of the 2Nd Workshop on Middleware for
  Context-Aware Applications in the IoT, M4IoT 2015, ACM, New York, NY, USA,
  2015, pp. 1--6.
\newblock \href {http://dx.doi.org/10.1145/2836127.2836132}
  {\path{doi:10.1145/2836127.2836132}}.
\newline\urlprefix\url{http://doi.acm.org.proxy.bnl.lu/10.1145/2836127.2836132}

\bibitem{bhatotia2014slider}
P.~Bhatotia, U.~A. Acar, F.~P. Junqueira, R.~Rodrigues, {Slider: incremental
  sliding window analytics}, in: Proceedings of the 15th International
  Middleware Conference, ACM, 2014, pp. 61--72.

\bibitem{Rusitschka:2013:AMR:2541596.2541601}
S.~Rusitschka, C.~Doblander, C.~Goebel, H.-A. Jacobsen,
  \href{http://doi.acm.org.proxy.bnl.lu/10.1145/2541596.2541601}{{Adaptive
  Middleware for Real-time Prescriptive Analytics in Large Scale Power
  Systems}}, in: Proceedings of the Industrial Track of the 13th
  ACM/IFIP/USENIX International Middleware Conference, Middleware Industry '13,
  ACM, New York, NY, USA, 2013, pp. 5:1--5:6.
\newblock \href {http://dx.doi.org/10.1145/2541596.2541601}
  {\path{doi:10.1145/2541596.2541601}}.
\newline\urlprefix\url{http://doi.acm.org.proxy.bnl.lu/10.1145/2541596.2541601}

\bibitem{HaasVLDB11}
P.~J. Haas, P.~P. Maglio, P.~G. Selinger, W.~C. Tan,
  \href{http://www.vldb.org/pvldb/vol4/p1486-haas.pdf}{{Data is Dead... Without
  What-If Models}}, {PVLDB} 4~(12) (2011) 1486--1489.
\newline\urlprefix\url{http://www.vldb.org/pvldb/vol4/p1486-haas.pdf}

\bibitem{Gonzalez:2012:PDG:2387880.2387883}
J.~E. Gonzalez, Y.~Low, H.~Gu, D.~Bickson, C.~Guestrin,
  \href{http://dl.acm.org/citation.cfm?id=2387880.2387883}{Powergraph:
  Distributed graph-parallel computation on natural graphs}, in: Proceedings of
  the 10th USENIX Conference on Operating Systems Design and Implementation,
  OSDI'12, USENIX Association, Berkeley, CA, USA, 2012, pp. 17--30.
\newline\urlprefix\url{http://dl.acm.org/citation.cfm?id=2387880.2387883}

\bibitem{Low:2012:DGF:2212351.2212354}
Y.~Low, D.~Bickson, J.~Gonzalez, C.~Guestrin, A.~Kyrola, J.~M. Hellerstein,
  \href{https://doi.org/10.14778/2212351.2212354}{Distributed graphlab: A
  framework for machine learning and data mining in the cloud}, Proc. VLDB
  Endow. 5~(8) (2012) 716--727.
\newblock \href {http://dx.doi.org/10.14778/2212351.2212354}
  {\path{doi:10.14778/2212351.2212354}}.
\newline\urlprefix\url{https://doi.org/10.14778/2212351.2212354}

\bibitem{Malewicz:2010:PSL:1807167.1807184}
G.~Malewicz, M.~H. Austern, A.~J. Bik, J.~C. Dehnert, I.~Horn, N.~Leiser,
  G.~Czajkowski, {Pregel: A System for Large-scale Graph Processing}, in:
  Proceedings of the 2010 ACM SIGMOD International Conference on Management of
  Data, SIGMOD '10, ACM, New York, NY, USA, 2010, pp. 135--146.

\bibitem{Leskovec:2005:GOT:1081870.1081893}
J.~Leskovec, J.~Kleinberg, C.~Faloutsos,
  \href{http://doi.acm.org/10.1145/1081870.1081893}{Graphs over time:
  Densification laws, shrinking diameters and possible explanations}, in:
  Proceedings of the Eleventh ACM SIGKDD International Conference on Knowledge
  Discovery in Data Mining, KDD '05, ACM, New York, NY, USA, 2005, pp.
  177--187.
\newblock \href {http://dx.doi.org/10.1145/1081870.1081893}
  {\path{doi:10.1145/1081870.1081893}}.
\newline\urlprefix\url{http://doi.acm.org/10.1145/1081870.1081893}

\bibitem{hartmannSeke14}
T.~Hartmann, F.~Fouquet, G.~Nain, B.~Morin, J.~Klein, Y.~Le~Traon, Reasoning at
  runtime using time-distorted contexts: Amodels@run.time based approach, in:
  (SEKE'14) 26th International Conference on Software Engineering and Knowledge
  Engineering, 2014.

\bibitem{Iyer:2016:TGP:2960414.2960419}
A.~P. Iyer, L.~E. Li, T.~Das, I.~Stoica,
  \href{http://doi.acm.org/10.1145/2960414.2960419}{{Time-evolving Graph
  Processing at Scale}}, in: Proceedings of the Fourth International Workshop
  on Graph Data Management Experiences and Systems, GRADES '16, ACM, New York,
  NY, USA, 2016, pp. 5:1--5:6.
\newblock \href {http://dx.doi.org/10.1145/2960414.2960419}
  {\path{doi:10.1145/2960414.2960419}}.
\newline\urlprefix\url{http://doi.acm.org/10.1145/2960414.2960419}

\bibitem{Han:2014:CGE:2592798.2592799}
W.~Han, Y.~Miao, K.~Li, M.~Wu, F.~Yang, L.~Zhou, V.~Prabhakaran, W.~Chen,
  E.~Chen, {Chronos: A Graph Engine for Temporal Graph Analysis}, in:
  Proceedings of the Ninth European Conference on Computer Systems, EuroSys
  '14, ACM, New York, NY, USA, 2014, pp. 1:1--1:14.

\bibitem{Miao:2015:ISS:2809503.2700302}
Y.~Miao, W.~Han, K.~Li, M.~Wu, F.~Yang, L.~Zhou, V.~Prabhakaran, E.~Chen,
  W.~Chen, \href{http://doi.acm.org/10.1145/2700302}{Immortalgraph: A system
  for storage and analysis of temporal graphs}, Trans. Storage 11~(3) (2015)
  14:1--14:34.
\newblock \href {http://dx.doi.org/10.1145/2700302}
  {\path{doi:10.1145/2700302}}.
\newline\urlprefix\url{http://doi.acm.org/10.1145/2700302}

\bibitem{Lin:2003:SRT:882082.882086}
J.~Lin, E.~Keogh, S.~Lonardi, B.~Chiu,
  \href{http://doi.acm.org/10.1145/882082.882086}{A symbolic representation of
  time series, with implications for streaming algorithms}, in: Proceedings of
  the 8th ACM SIGMOD Workshop on Research Issues in Data Mining and Knowledge
  Discovery, DMKD '03, ACM, New York, NY, USA, 2003, pp. 2--11.
\newblock \href {http://dx.doi.org/10.1145/882082.882086}
  {\path{doi:10.1145/882082.882086}}.
\newline\urlprefix\url{http://doi.acm.org/10.1145/882082.882086}

\bibitem{everett1957relative}
H.~Everett, \href{http://link.aps.org/doi/10.1103/RevModPhys.29.454}{"relative
  state" formulation of quantum mechanics}, Rev. Mod. Phys. 29 (1957) 454--462.
\newblock \href {http://dx.doi.org/10.1103/RevModPhys.29.454}
  {\path{doi:10.1103/RevModPhys.29.454}}.
\newline\urlprefix\url{http://link.aps.org/doi/10.1103/RevModPhys.29.454}

\bibitem{low2014graphlab}
Y.~Low, J.~E. Gonzalez, A.~Kyrola, D.~Bickson, C.~E. Guestrin, J.~Hellerstein,
  {Graphlab: A new framework for parallel machine learning}, arXiv preprint
  arXiv:1408.2041.

\bibitem{xin2013graphx}
R.~S. Xin, J.~E. Gonzalez, M.~J. Franklin, I.~Stoica, Graphx: A resilient
  distributed graph system on spark, in: First International Workshop on Graph
  Data Management Experiences and Systems, GRADES '13, ACM, New York, NY, USA,
  2013, pp. 2:1--2:6.

\bibitem{miller2013graph}
J.~J. Miller, {Graph Database Applications and Concepts with Neo4j}, in:
  Proceedings of the Southern Association for Information Systems Conference,
  Atlanta, GA, USA March 23rd-24th, 2013.

\bibitem{Bahmani:2012:PEG:2339530.2339539}
B.~Bahmani, R.~Kumar, M.~Mahdian, E.~Upfal, {PageRank on an Evolving Graph},
  in: Proceedings of the 18th ACM SIGKDD International Conference on Knowledge
  Discovery and Data Mining, KDD '12, ACM, New York, NY, USA, 2012, pp. 24--32.

\bibitem{Cattuto:2013:TSN:2484425.2484442}
C.~Cattuto, M.~Quaggiotto, A.~Panisson, A.~Averbuch, {Time-varying Social
  Networks in a Graph Database: A Neo4J Use Case}, in: First International
  Workshop on Graph Data Management Experiences and Systems, GRADES '13, ACM,
  New York, NY, USA, 2013, pp. 11:1--11:6.

\bibitem{khurana2015storing}
U.~Khurana, A.~Deshpande, Storing and analyzing historical graph data at scale,
  arXiv preprint arXiv:1509.08960.

\bibitem{schrodinger1935present}
E.~Schr{\"o}dinger, The present status of quantum mechanics, Die
  Naturwissenschaften 23~(48) (1935) 1--26.

\bibitem{1507024}
S.~M. Amin, B.~F. Wollenberg, {Toward a smart grid: power delivery for the 21st
  century}, IEEE Power and Energy Magazine 3~(5) (2005) 34--41.
\newblock \href {http://dx.doi.org/10.1109/MPAE.2005.1507024}
  {\path{doi:10.1109/MPAE.2005.1507024}}.

\bibitem{5357331}
H.~Farhangi, {The path of the smart grid}, IEEE Power and Energy Magazine 8~(1)
  (2010) 18--28.
\newblock \href {http://dx.doi.org/10.1109/MPE.2009.934876}
  {\path{doi:10.1109/MPE.2009.934876}}.

\bibitem{DBLP:conf/smartgridcomm/0001MFRMKT15}
T.~Hartmann, A.~Moawad, F.~Fouquet, Y.~Reckinger, T.~Mouelhi, J.~Klein, Y.~L.
  Traon,
  \href{http://dx.doi.org/10.1109/SmartGridComm.2015.7436414}{{Suspicious
  electric consumption detection based on multi-profiling using live machine
  learning}}, in: 2015 {IEEE} International Conference on Smart Grid
  Communications, SmartGridComm 2015, Miami, FL, USA, November 2-5, 2015,
  {IEEE}, 2015, pp. 891--896.
\newblock \href {http://dx.doi.org/10.1109/SmartGridComm.2015.7436414}
  {\path{doi:10.1109/SmartGridComm.2015.7436414}}.
\newline\urlprefix\url{http://dx.doi.org/10.1109/SmartGridComm.2015.7436414}

\bibitem{clifford1983formal}
J.~Clifford, D.~S. Warren, Formal semantics for time in databases, ACM
  Transactions on Database Systems (TODS) 8~(2) (1983) 214--254.

\bibitem{segev1987logical}
A.~Segev, A.~Shoshani, Logical modeling of temporal data, in: ACM Sigmod
  Record, Vol.~16, ACM, 1987, pp. 454--466.

\bibitem{influxdb}
{influxdb - TIME-SERIES DATA STORAGE DOCS}. [{Online}], Available:
  \url{https://influxdata.com/time-series-platform/influxdb}.

\bibitem{neo4j}
{neo4j}. [{Online}], Available: \url{http://neo4j.com}.

\bibitem{Cheng:2012:KTP:2168836.2168846}
R.~Cheng, J.~Hong, A.~Kyrola, Y.~Miao, X.~Weng, M.~Wu, F.~Yang, L.~Zhou,
  F.~Zhao, E.~Chen,
  \href{http://doi.acm.org/10.1145/2168836.2168846}{{Kineograph: Taking the
  Pulse of a Fast-changing and Connected World}}, in: Proceedings of the 7th
  ACM European Conference on Computer Systems, EuroSys '12, ACM, New York, NY,
  USA, 2012, pp. 85--98.
\newblock \href {http://dx.doi.org/10.1145/2168836.2168846}
  {\path{doi:10.1145/2168836.2168846}}.
\newline\urlprefix\url{http://doi.acm.org/10.1145/2168836.2168846}

\bibitem{Fabrega95copyon}
F.~J.~T. Fábrega, F.~Javier, J.~D. Guttman, {Copy on Write} (1995).

\bibitem{Guibas:1978:DFB:1382432.1382565}
L.~J. Guibas, R.~Sedgewick, \href{http://dx.doi.org/10.1109/SFCS.1978.3}{A
  dichromatic framework for balanced trees}, in: Proceedings of the 19th Annual
  Symposium on Foundations of Computer Science, SFCS '78, IEEE Computer
  Society, Washington, DC, USA, 1978, pp. 8--21.
\newblock \href {http://dx.doi.org/10.1109/SFCS.1978.3}
  {\path{doi:10.1109/SFCS.1978.3}}.
\newline\urlprefix\url{http://dx.doi.org/10.1109/SFCS.1978.3}

\bibitem{Veiga:2015:ACG:2814576.2814813}
L.~Veiga, R.~Bruno, P.~Ferreira,
  \href{http://doi.acm.org.proxy.bnl.lu/10.1145/2814576.2814813}{{Asynchronous
  Complete Garbage Collection for Graph Data Stores}}, in: Proceedings of the
  16th Annual Middleware Conference, Middleware '15, ACM, New York, NY, USA,
  2015, pp. 112--124.
\newblock \href {http://dx.doi.org/10.1145/2814576.2814813}
  {\path{doi:10.1145/2814576.2814813}}.
\newline\urlprefix\url{http://doi.acm.org.proxy.bnl.lu/10.1145/2814576.2814813}

\bibitem{conf/usenix/ChenZL03}
Z.~Chen, Y.~Zhou, K.~Li, Eviction-based cache placement for storage caches.,
  in: USENIX Annual Technical Conference, General Track, USENIX, 2003, pp.
  269--281.

\bibitem{rocksdb}
{RocksDB}. [{Online}], Available: \url{http://rocksdb.org}.

\bibitem{LevelDB}
Leveldb, \url{http://leveldb.org}, accessed: 2015-03-02.

\bibitem{cassandra}
{Cassandra}. [{Online}], Available: \url{http://cassandra.apache.org/}.

\bibitem{george2011hbase}
L.~George, HBase: the definitive guide, " O'Reilly Media, Inc.", 2011.

\bibitem{hpc}
{HPC\@Uni.lu}. [{Online}], Available: \url{https://hpc.uni.lu}.

\bibitem{mwg_experiments}
{GreyCat Experiments}. [{Online}], Available:
  \url{http://datathings.com/experiments/greycat-mwg-benchmark/}.

\bibitem{london_data_set}
{London DataStore}. [{Online}], Available:
  \url{http://data.london.gov.uk/dataset/smartmeter-energy-use-data-in-london-households}.

\bibitem{DBLP:conf/smartgridcomm/0001FKTPTR14}
T.~Hartmann, F.~Fouquet, J.~Klein, Y.~L. Traon, A.~Pelov, L.~Toutain,
  T.~Ropitault,
  \href{http://dx.doi.org/10.1109/SmartGridComm.2014.7007684}{{Generating
  realistic Smart Grid communication topologies based on real-data}}, in: 2014
  {IEEE} International Conference on Smart Grid Communications, SmartGridComm
  2014, Venice, Italy, November 3-6, 2014, {IEEE}, 2014, pp. 428--433.
\newblock \href {http://dx.doi.org/10.1109/SmartGridComm.2014.7007684}
  {\path{doi:10.1109/SmartGridComm.2014.7007684}}.
\newline\urlprefix\url{http://dx.doi.org/10.1109/SmartGridComm.2014.7007684}

\bibitem{graphdb_bench}
{graphdb-benchmarks}. [{Online}], Available:
  \url{https://github.com/socialsensor/graphdb-benchmarks}.

\bibitem{Beis2015}
S.~Beis, S.~Papadopoulos, Y.~Kompatsiaris, {Benchmarking graph databases on the
  problem of community detection}, in: New Trends in Database and Information
  Systems II, Springer, 2015, pp. 3--14.

\bibitem{stanford_dataset}
{Stanford Large Network Dataset Collection}. [{Online}], Available:
  \url{http://snap.stanford.edu/data}.

\bibitem{DBLP:journals/corr/abs-1205-6233}
J.~Yang, J.~Leskovec, \href{http://arxiv.org/abs/1205.6233}{{Defining and
  Evaluating Network Communities based on Ground-truth}}, CoRR abs/1205.6233.
\newline\urlprefix\url{http://arxiv.org/abs/1205.6233}

\bibitem{influx_bench}
{influxDB}. [{Online}], Available: \url{http://tinyurl.com/influxblog}.

\bibitem{srinivas1994genetic}
M.~Srinivas, L.~M. Patnaik, Genetic algorithms: A survey, Computer 27~(6)
  (1994) 17--26.

\bibitem{codd1993}
E.~F. Codd, S.~B. Codd, C.~T. Salley, {Providing OLAP (On-Line Analytical
  Processing) to User-Analysts: An IT Mandate}, E. F. Codd and Associates
  (1993).

\bibitem{cohen09madskills}
J.~Cohen, B.~Dolan, M.~Dunlap, J.~M. Hellerstein, C.~Welton,
  \href{http://dl.acm.org/citation.cfm?id=1687553.1687576}{Mad skills: new
  analysis practices for big data}, Proceedings VLDB Endowment 2~(2) (2009)
  1481--1492.
\newline\urlprefix\url{http://dl.acm.org/citation.cfm?id=1687553.1687576}

\bibitem{hadoop}
{Apache Hadoop}. [{Online}], Available: \url{https://hadoop.apache.org}.

\bibitem{Kulkarni:2015:THS:2723372.2742788}
S.~Kulkarni, N.~Bhagat, M.~Fu, V.~Kedigehalli, C.~Kellogg, S.~Mittal, J.~M.
  Patel, K.~Ramasamy, S.~Taneja, {Twitter Heron: Stream Processing at Scale},
  in: Proceedings of the 2015 ACM SIGMOD International Conference on Management
  of Data, SIGMOD'15, ACM, New York, NY, USA, 2015, pp. 239--250.

\bibitem{zaharia2012resilient}
M.~Zaharia, M.~Chowdhury, T.~Das, A.~Dave, J.~Ma, M.~McCauley, M.~J. Franklin,
  S.~Shenker, I.~Stoica, Resilient distributed datasets: A fault-tolerant
  abstraction for in-memory cluster computing, in: Proceedings of the 9th
  USENIX conference on Networked Systems Design and Implementation, USENIX
  Association, 2012, pp. 2--2.

\bibitem{Canas:2015:GGP:2814576.2814812}
C.~Ca\~{n}as, E.~Pacheco, B.~Kemme, J.~Kienzle, H.-A. Jacobsen,
  \href{http://doi.acm.org.proxy.bnl.lu/10.1145/2814576.2814812}{{GraPS: A
  Graph Publish/Subscribe Middleware}}, in: Proceedings of the 16th Annual
  Middleware Conference, Middleware '15, ACM, New York, NY, USA, 2015, pp.
  1--12.
\newblock \href {http://dx.doi.org/10.1145/2814576.2814812}
  {\path{doi:10.1145/2814576.2814812}}.
\newline\urlprefix\url{http://doi.acm.org.proxy.bnl.lu/10.1145/2814576.2814812}

\bibitem{Kermarrec:2015:SOL:2814576.2814810}
A.-M. Kermarrec, F.~Taiani, J.~M. Tirado,
  \href{http://doi.acm.org.proxy.bnl.lu/10.1145/2814576.2814810}{{Scaling Out
  Link Prediction with SNAPLE: 1 Billion Edges and Beyond}}, in: Proceedings of
  the 16th Annual Middleware Conference, Middleware '15, ACM, New York, NY,
  USA, 2015, pp. 247--258.
\newblock \href {http://dx.doi.org/10.1145/2814576.2814810}
  {\path{doi:10.1145/2814576.2814810}}.
\newline\urlprefix\url{http://doi.acm.org.proxy.bnl.lu/10.1145/2814576.2814810}

\bibitem{ching2015one}
A.~Ching, S.~Edunov, M.~Kabiljo, D.~Logothetis, S.~Muthukrishnan, {One trillion
  edges: graph processing at Facebook-scale}, Proceedings of the VLDB Endowment
  8~(12) (2015) 1804--1815.

\bibitem{Roy:2015:CSG:2815400.2815408}
A.~Roy, L.~Bindschaedler, J.~Malicevic, W.~Zwaenepoel, {Chaos: Scale-out Graph
  Processing from Secondary Storage}, in: Proceedings of the 25th Symposium on
  Operating Systems Principles, SOSP '15, ACM, New York, NY, USA, 2015, pp.
  410--424.

\bibitem{Shao:2013:TDG:2463676.2467799}
B.~Shao, H.~Wang, Y.~Li,
  \href{http://doi.acm.org/10.1145/2463676.2467799}{{Trinity: A Distributed
  Graph Engine on a Memory Cloud}}, in: Proceedings of the 2013 ACM SIGMOD
  International Conference on Management of Data, SIGMOD '13, ACM, New York,
  NY, USA, 2013, pp. 505--516.
\newblock \href {http://dx.doi.org/10.1145/2463676.2467799}
  {\path{doi:10.1145/2463676.2467799}}.
\newline\urlprefix\url{http://doi.acm.org/10.1145/2463676.2467799}

\bibitem{ariav1986temporally}
G.~Ariav, A temporally oriented data model, ACM Transactions on Database
  Systems (TODS) 11~(4) (1986) 499--527.

\bibitem{Salzberg:1999:CAM:319806.319816}
B.~Salzberg, V.~J. Tsotras, {Comparison of Access Methods for Time-evolving
  Data}, ACM Comput. Surv. 31~(2) (1999) 158--221.

\bibitem{chang2008bigtable}
F.~Chang, J.~Dean, S.~Ghemawat, W.~C. Hsieh, D.~A. Wallach, M.~Burrows,
  T.~Chandra, A.~Fikes, R.~E. Gruber, {Bigtable: A Distributed Storage System
  for Structured Data}, ACM Trans. Comput. Syst. 26~(2) (2008) 4:1--4:26.

\bibitem{atlas}
{Netflix Atlas}. [{Online}], Available:
  \url{https://github.com/Netflix/atlas/wiki/Graph}.

\bibitem{open_tbsd}
{OpenTSDB}. [{Online}], Available: \url{http://opentsdb.net}.

\bibitem{rdd_tool}
{RDDTool}. [{Online}], Available: \url{http://oss.oetiker.ch/rrdtool}.

\bibitem{wu2014path}
H.~Wu, J.~Cheng, S.~Huang, Y.~Ke, Y.~Lu, Y.~Xu, Path problems in temporal
  graphs, Proceedings of the VLDB Endowment 7~(9) (2014) 721--732.

\bibitem{grandi2010t}
F.~Grandi, {T-SPARQL: A TSQL2-like Temporal Query Language for RDF}, in: ADBIS
  (Local Proceedings), Citeseer, 2010, pp. 21--30.

\bibitem{perez2009semantics}
J.~P{\'e}rez, M.~Arenas, C.~Gutierrez, {Semantics and Complexity of SPARQL},
  ACM Trans. Database Syst. 34~(3) (2009) 16:1--16:45.

\bibitem{Balmin:2000:HQO:645926.672016}
A.~Balmin, T.~Papadimitriou, Y.~Papakonstantinou,
  \href{http://dl.acm.org/citation.cfm?id=645926.672016}{Hypothetical queries
  in an olap environment}, in: Proceedings of the 26th International Conference
  on Very Large Data Bases, VLDB '00, Morgan Kaufmann Publishers Inc., San
  Francisco, CA, USA, 2000, pp. 220--231.
\newline\urlprefix\url{http://dl.acm.org/citation.cfm?id=645926.672016}

\bibitem{Griffin:1997:FIH:253260.253304}
T.~Griffin, R.~Hull,
  \href{http://doi.acm.org.proxy.bnl.lu/10.1145/253260.253304}{A framework for
  implementing hypothetical queries}, in: Proceedings of the 1997 ACM SIGMOD
  International Conference on Management of Data, SIGMOD '97, ACM, New York,
  NY, USA, 1997, pp. 231--242.
\newblock \href {http://dx.doi.org/10.1145/253260.253304}
  {\path{doi:10.1145/253260.253304}}.
\newline\urlprefix\url{http://doi.acm.org.proxy.bnl.lu/10.1145/253260.253304}

\bibitem{arenas1998hypothetical}
M.~Arenas, L.~Bertossi, Hypothetical temporal queries in databases, in:
  Proceedings “ACM SIGMOD/PODS 5th International Workshop on Knowledge
  Representation meets Databases (KRDB’98): Innovative Application
  Programming and Query Interfaces, Citeseer, 1998.

\bibitem{hypo_temporal_reasoning}
M.~Arenas, L.~Bertossi,
  \href{http://dx.doi.org/10.1023/A:1016524013831}{{Hypothetical Temporal
  Reasoning in Databases}}, Journal of Intelligent Information Systems 19~(2)
  231--259.
\newblock \href {http://dx.doi.org/10.1023/A:1016524013831}
  {\path{doi:10.1023/A:1016524013831}}.
\newline\urlprefix\url{http://dx.doi.org/10.1023/A:1016524013831}

\bibitem{DBLP:conf/middleware/KermarrecTT15}
A.~Kermarrec, F.~Ta{\"{\i}}ani, J.~M. Tirado, {Scaling Out Link Prediction with
  SNAPLE: 1 Billion Edges and Beyond}, in: Middleware, {ACM}, 2015, pp.
  247--258.

\end{thebibliography}

\end{document}